\documentclass[twocolumn,notitlepage,prl,superscriptaddress,longtable, longbibliography]{revtex4-1}
\usepackage{mhchem}
\usepackage{amsthm}
\usepackage{amsmath}
\usepackage{amssymb}
\usepackage{mathdots}
\usepackage{graphicx}
\usepackage{mathrsfs}
\usepackage{longtable}
\usepackage{multirow}
\makeatletter
\@ifundefined{textcolor}{}
{%
 \definecolor{BLACK}{gray}{0}
 \definecolor{WHITE}{gray}{1}
 \definecolor{RED}{rgb}{1,0,0}
 \definecolor{GREEN}{rgb}{0,1,0}
 \definecolor{BLUE}{rgb}{0,0,1}
 \definecolor{CYAN}{cmyk}{1,0,0,0}
 \definecolor{MAGENTA}{cmyk}{0,1,0,0}
 \definecolor{YELLOW}{cmyk}{0,0,1,0}
}

\usepackage{braket}
\usepackage[backref=true,
bookmarksnumbered=true,
bookmarks=true,
bookmarksopen=true,
colorlinks=true,
citecolor=blue,
linkcolor=blue,
anchorcolor=green,
urlcolor=blue,unicode=false]{hyperref}

\renewcommand{\v}[1]{\ensuremath{\mathbf{#1}}} % for vectors

\let\baraccent=\= % rename builtin command \= to \baraccent
\renewcommand{\=}[1]{\stackrel{#1}{=}} % for putting numbers above =

\DeclareMathOperator{\Tr}{Tr}

\begin{document}
\title{Distinct Floquet topological classifications from color-decorated frequency lattices with space-time symmetries}

\author{Ilyoun Na}
    \affiliation{Department of Physics, University of California, Berkeley, California 94720, USA}
    \affiliation{Materials Sciences Division, Lawrence Berkeley National Laboratory, Berkeley, California 94720, USA}
\author{Jack Kemp}
    \affiliation{Department of Physics, Harvard University, Cambridge, Massachusetts 02138 USA}
\author{Robert-Jan Slager}
    \affiliation{TCM Group, Cavendish Laboratory, Department of Physics, J J Thomson Avenue, Cambridge CB3 0HE, United Kingdom}
\author{Yang Peng}
    \email{Corresponding author: yang.peng@csun.edu}
    \affiliation{Department of Physics and Astronomy, California State University, Northridge, Northridge, California 91330, USA}
    \affiliation{Department of Physics, California Institute of Technology, Pasadena, California 91125, USA}

\date{\today}

\begin{abstract}
We consider nontrivial topological phases in Floquet systems using unitary loops and stroboscopic evolutions under a static Floquet Hamiltonian $H_F$ in the presence of dynamical space-time symmetries $G$. While the latter has been subject of out-of-equilibrium classifications that extend the ten-fold way and systems with additional crystalline symmetries to periodically driven systems, we explore the anomalous topological zero modes that arise in $H_F$ from the coexistence of a dynamical space-time symmetry $M$ and antisymmetry $A$ of $G$, and classify them using a frequency-domain formulation. Moreover, we provide an interpretation of the resulting Floquet topological phases using a frequency lattice with a decoration represented by color degrees of freedom on the lattice vertices. These colors correspond to the coefficient $N$ of the group extension $\Tilde{G}$ of $G$ along the frequency lattice, given by $N=Z\rtimes H^1[A,M]$. The distinct topological classifications that arise at different energy gaps in its quasi-energy spectrum are described by the torsion product of the cohomology group $H^{2}[G,N]$ classifying the group extension.
\end{abstract}

\maketitle

Remarkable progress has recently been made towards understanding the interplay between symmetry and topology. Equilibrium topological states of matter have been systematically classified by their symmetries, including local symmetries such as time-reversal, particle-hole, and chiral~\cite{Schnyder2008,Kitaev2009,Ryu2010,Teo2010,Chiu2016}, as well as symmorphic and non-symmorphic crystalline symmetries~\cite{Fu2011,Chiu2013,Slager2013,Shiozaki2014, Ando2015,Slager2015,Shiozaki2016,Kruthoff2017,Bradlyn2017,Song2018,Song2019,Po2020,Bouhon2020}. Moving beyond equilibrium, periodically-driven Floquet systems supporting exotic new phases have been discovered with no equilibrium analogues ~\cite{Kitagawa2011,Rudner2013,Nathan2015,Keyserlingk2016_1,Keyserlingk2016_2, Else2016,Potter2016,Khemani2016,Unal2019,Unal2020,Vu2021,Slager2022}, including the anomalous Floquet insulator~\cite{Kitagawa2010,Rudner2013,Titum2016,Nathan2017,Nathan2019,Wintersperger2020}. For the Altland-Zirnbauer (AZ) symmetry classes, a complete classification of non-interacting Floquet topological insulators and superconductors has been obtained~\cite{Roy2017,Yao2017}, which is further enriched when considering crystalline symmetries~\cite{Fulga2019,Yu2021}.

Intuitively, the distinction between the static and Floquet classifications for these symmetry classes arises because there is no notion of the Fermi energy in Floquet systems. This means that we must classify all $n$ gaps in the quasi-energy spectrum, as opposed to simply the gap at the Fermi energy. Thus, the Abelian group $\mathcal{G}\in \{\mathbb{Z}_1,\mathbb{Z}_2, \mathbb{Z}\}$ classifying the phases in the static case generalizes to $\mathcal{G}^{\times n}$ in the Floquet~\cite{Roy2017}.

%to distinguish where this difference in classification arises at the level of the unitary

In order to isolate the phases that do not exist in the static limit, one can decompose the Floquet unitary time-evolution operator as $\hat{U}(t)=\hat{V}(t)e^{-i\hat{H}_Ft}$~\cite{Goldman2014}. The time-independent Floquet Hamiltonian $\hat{H}_F$ governs the stroboscopic time evolution, while the time-dependent unitary loop $\hat{V}(t)=\hat{V}(t+T)$ inherits the Floquet time period $T$. As the time-periodic terms in the Hamiltonian are turned off adiabatically, $\hat{H}_F$ approaches the static Hamiltonian, leading one to expect that any nontrivial topological phases induced by $\hat{H}_F$ would be equivalent to those in static systems. Accordingly, the literature on Floquet topological phases has focused mainly on the topology of the loop $\hat{V}(t)$. 

In the AZ and crystalline symmetry classes, $\hat{V}(t)$ is indeed solely responsible for the anomalous Floquet topological modes~\cite{Rudner2013}. But recently, intertwined nonsymmorphic space-time symmetries~\cite{Morimoto2017,Xu2018} have gained interest, which combine temporal translation with spatial transformations. It has been shown that these space-time symmetries lead to a richer topological classification compared to that obtained using purely static symmetries~\cite{Peng2019,Peng2020, Swati2020, Peng2022}. In particular, in Ref.~\cite{Peng2020}, one of the present authors classified the Floquet systems with order-two space-time symmetries/antisymmetries using unitary loop in the case of spectral flattened Floquet Hamiltonian, $\hat{H}_F=0$. 

In this Letter, we introduce a class of anomalous Floquet topological zero modes that arise directly from $\hat{H}_F$ rather than the unitary loop, as a consequence of the combination of an order-two space-time symmetry with a static antisymmetry. We classify these modes using a frequency-domain formulation and find that the Abelian group classifying the phases does not have the usual $\mathcal{G}^{\times n}$ structure. Instead, distinct topological classifications may appear at different energy gaps in the quasi-energy spectrum. 
We provide a clear interpretation of these new topological phases by explaining how the action of the two symmetries decorates the frequency lattice with an emergent `color' degree of freedom. In particular, we show that on the frequency lattice, the order-two space-time symmetry is on-site, but alternates in sign between even and odd sites. This distinction is typically hidden by the equivalence of even and odd lattice sites, but is revealed by the non-local action of the antisymmetry operation. 
We moreover demonstrate that the equivalence classes of the enlarged symmetry group $\Tilde{G}$ of the decorated frequency lattice are fully characterized by the cohomology group $H^2[G,N]$, where the coefficient $N$ captures the decoration. Additionally, we show how the torsion product of $H^2[G,N]$ naturally accounts for the presence of these distinct topological classifications.  

\emph{Symmetries and Frequency-domain formulation}---We study non-interacting Floquet systems which exhibit an order-two space-time unitary symmetry (such as time-glide~\cite{Morimoto2017}) and a static antiunitary antisymmetry. In momentum-space representation, the Bloch Hamiltonian $H(\mathit{\mathbf{k}},t)$, transforms respectively under these as 
\begin{align}
    U_{T/2}H(\mathit{\mathbf{k}},t)U_{T/2}^{\dag} =& H(-\mathit{\mathbf{k_{\parallel}}},\mathit{\mathbf{k_{\perp}}},t+T/2), \label{eq:Noninteracting Hamiltonian with coexisting symmetries} \\
    \bar{A}_{0}H(\mathit{\mathbf{k}},t)\bar{A}_{0}^{-1} =& -H(\mathit{\mathbf{k_{\parallel}}},-\mathit{\mathbf{k_{\perp}}},-t), \nonumber    
\end{align}
where $\mathit{\mathbf{k_{\parallel}}}$ and $\mathit{\mathbf{k_{\perp}}}$ denote the components of the momentum which flip under the symmetry/antisymmetry respectively~\cite{Peng2020,SM}.

Recall that a Floquet system can be described by a static frequency-enlarged Hamiltonian $\mathcal{H}(\mathit{\mathbf{k}})$ whose matrix blocks are given by $\mathcal{H}_{m,n}(\mathit{\mathbf{k}})=-m\Omega\cdot\delta_{m,n}\mathbb{I}+h_{m-n}(\mathit{\mathbf{k}})$ where $h_{m-n}(\mathit{\mathbf{k}})=(1/T)\int_{0}^{T}dtH(\mathit{\mathbf{k}},t)e^{i(m-n)\Omega t}=h_{n-m}(\mathit{\mathbf{k}})^{\dag}$. Here, $\mathbb{I}$ is the identity matrix of the same size as $H(\mathit{\mathbf{k}})$ and $m,n\in\mathbb{Z}$ are frequency indices~\cite{Rudner2013}. We now consider the action of the symmetries in Eq.~\eqref{eq:Noninteracting Hamiltonian with coexisting symmetries} on this static enlarged Hamiltonian. 

One can readily show that the enlarged Hamiltonian $\mathcal{H}(\mathit{\mathbf{k}})$ inherits a spatial unitary symmetry $\mathcal{U}$ and antiunitary antisymmetry $\mathcal{\bar{A}}$ from the original space-time symmetry $U_{T/2}$ and antisymmetry $\bar{A}_0$ in $H(\mathit{\mathbf{k}},t)$. Indeed, given $U_{T/2}h_n(\mathit{\mathbf{k}})U^{\dag}_{T/2}=(-1)^{n}h_n(-\mathit{\mathbf{k_{\parallel}}},\mathit{\mathbf{k}_{\perp}})$ and $\bar{A}_{0}h_n(\mathit{\mathbf{k}})\bar{A}^{-1}_{0}=-h_{-n}(\mathit{\mathbf{k_{\parallel}}},-\mathit{\mathbf{k}_{\perp}})$, it immediately follows
\begin{equation} \label{eq:Unitary,Antiunitary Syms in Enlarged Hamiltonian}
    \mathcal{U}=\begin{pmatrix}
                \ddots &  &  & \\ 
                & U_{T/2} &  & \\ 
                &  & -U_{T/2} & \\ 
                &  &  & \ddots  \\ 
                \end{pmatrix},~ 
    \mathcal{\bar{A}}=\begin{pmatrix}
                &  &  &  \iddots \\ 
                &  & \bar{A}_0 & \\ 
                & \bar{A}_0 &  & \\ 
                \iddots &  &  & \\ 
                \end{pmatrix},
\end{equation}
where the enlarged Hamiltonian satisfies $\mathcal{U}\mathcal{H}(\mathit{\mathbf{k}})\mathcal{U}^{\dag}=\mathcal{H}(-\mathit{\mathbf{k_{\parallel}}},\mathit{\mathbf{k_{\perp}}})$ and   $\mathcal{\bar{A}}\mathcal{H}(\mathit{\mathbf{k}})\mathcal{\bar{A}}^{-1}=-\mathcal{H}(\mathit{\mathbf{k_{\parallel}}},-\mathit{\mathbf{k_{\perp}}})$~\cite{Peng2020,SM}. 

Here, we identify a topological classification of nontrivial edge states appearing at the gap $E_F=0$ or $\Omega/2$ from that of truncated matrices made up of a finite number $2n+1$ or $2n$ matrix blocks respectively, under the assumption that the infinite-dimensional $\mathcal{H}$ will be recovered by taking the limit $n\rightarrow\infty$. The diagonal blocks of these truncated matrices run from $h_0-n\Omega$ to $h_0+n\Omega$ (or $n-1$ in the even block-dimension case). To gain intuition on the two distinct classification results at the two different gaps, we consider the $3\times3$ and $2\times2$ blocks of the truncated $\mathcal{H}$, 
\begin{align} \label{eq:Even/odd truncation of H_enlarged}
    \mathcal{H}_{odd}&=\begin{pmatrix}
                       h_0-\Omega & h^{\dag}_1 & h^{\dag}_2\\ 
                       h_1 & h_0 & h^{\dag}_1\\ 
                       h_2 & h_1 & h_0+\Omega
                      \end{pmatrix}, \nonumber \\ 
    \mathcal{H}_{even}&=\begin{pmatrix}
                       h_0-\Omega/2 & h^{\dag}_1 \\ 
                       h_1 & h_0+\Omega/2 \\ 
                       \end{pmatrix}-\frac{\Omega}{2}\rho_0,    
\end{align}
where $\rho_0$ is the identity in the two-Floquet-zone basis. The topological classification of the static $\mathcal{H}_{odd}$ and the first term of $\mathcal{H}_{even}$ in Eq.~\eqref{eq:Even/odd truncation of H_enlarged} characterizes the zero and $\pi$ modes at $E_F=0$ or $\Omega/2$ respectively. This can be generalized to any odd- or even- dimension block size. 

\emph{Model}---As a concrete example, we introduce a model of spinless fermions hopping on a bipartite one-dimensional lattice, which hosts anomalous Floquet topological zero modes protected by both a time-glide $U_{T/2}$ and an antisymmetry $\bar{A}_{0}$. The Hamiltonian is a four-step drive 
\begin{equation} \label{eq:four-step driven model}
    H(\mathit{k},t)=\begin{cases}
         H_1(\mathit{k})&(0 \leq t < \frac{T}{4}), \\ 
         H_2(\mathit{k})&(\frac{T}{4} \leq t < \frac{T}{2}), \\ 
         H_3(\mathit{k})&(\frac{T}{2} \leq t < \frac{3T}{4}),\\ 
         H_4(\mathit{k})&(\frac{3T}{4} \leq t < T), 
         \end{cases}
\end{equation}
where $H_1(k)=J_1\sigma_x+[\delta-2J_4\sin(2k)]\sigma_z$ and $H_2(k)=-J_2\sin(2k)\sigma_x-J_2\cos(2k)\sigma_y-2J_3\sin(2k)\sigma_z$. Here, the Pauli matrices $\sigma_i$ act on the sublattice degrees of freedom. This Hamiltonian satisfies $U_{T/2}H_{1,2}(k)U^{\dag}_{T/2}=H_{3,4}(-k)$, $U_{T/2}=\sigma_x$ and $\bar{A}_{0}H(k,t)\bar{A}^{-1}_{0}=-H(k,t)$, $\bar{A}_0=i\sigma_yK$ where $K$ is the complex conjugation. Here, $U_{T/2}^2=\mathbb{I}$, $\bar{A}_0^2=-\mathbb{I}$, and $\left \{U_{T/2},\bar{A}_0\right \}=0$. Its quasi-energy spectrum hosts a bulk gap around $E_F=0$, where nontrivial edge states appear~\cite{SM}. However, we show that its loop operator $\hat{V}(t)$ must be topologically trivial, by Hermitian mapping it to a two-dimensional static Hamiltonian $H_{\textrm{eff}}(k,t)$ of class AIII~\cite{Roy2017,Morimoto2017,SM}. This 2D static Hamiltonian does not support nontrivial topological phases from the K-theory classification~\cite{Shiozaki2014,SM}, as the corresponding K group is $\mathbb{Z}_1\equiv 0$. The existence of non-trivial edge states is thus quite surprising, and suggests that the classification of Floquet systems with dynamical space-time symmetries/antisymmetries may not be the familiar $\mathcal{G}^{\times n}$ from the AZ and crystalline symmetries~\cite{Roy2017}. In the following, we employ the frequency-domain formulation to characterize the topological properties of these anomalous zero modes.

\emph{Topological invariants of anomalous topological zero modes}---The Hamiltonian $H(k,t)$ defined in Eq.~\eqref{eq:four-step driven model} maps to an enlarged $\mathcal{H}$ with $\mathcal{U}$ and $\mathcal{\bar{A}}$ satisfying $\mathcal{U}\mathcal{H}(k)\mathcal{U}^{\dag}=\mathcal{H}(-k)$, $\mathcal{U}^2=\mathbb{I}$ and $\mathcal{\bar{A}}\mathcal{H}(k)\mathcal{\bar{A}}^{-1}=-\mathcal{H}(k)$, $\mathcal{\bar{A}}^2=-\mathbb{I}$. $\mathcal{U}$ and $\mathcal{\bar{A}}$ anti-commute with $\mathcal{H}_{odd}$, but commute with $\mathcal{H}_{even}$ as defined in Eq.~\eqref{eq:Even/odd truncation of H_enlarged}. Defining a particle-hole symmetry as $\mathcal{C}=\mathcal{U}\mathcal{\bar{A}}$, $\mathcal{C}\mathcal{H}(k)\mathcal{C}^{-1}=-\mathcal{H}(-k)$, we find that the symmetry class of $\mathcal{H}_{odd}(\mathcal{H}_{even})$ is the 1D class D(C) with an effective reflection symmetry $\mathcal{U}$ anti-commuting(commuting) with $\mathcal{C}$ satisfying $\mathcal{C}^2=\mathbb{I}(-\mathbb{I})$ for $\mathcal{H}_{odd}(\mathcal{H}_{even})$. These different symmetry classes generate correspondingly different topological classifications at the zero($\pi$) gaps as $\mathbb{Z}$($\mathbb{Z}_1$)~\cite{Chiu2013,Shiozaki2014}, the latter matching the topologically trivial $\pi$ gap obtained from the loop $\hat{V}(t)$. The opposite commutation relations between $\mathcal{U}$ and $\mathcal{A}$ which account for this difference in classification originate from the order-two nature of the dynamical space-time symmetry that determines the factor $(-1)^n$ in $U_{T/2}h_n(k)U^{\dag}_{T/2}=(-1)^n h_n(-k)$. We call these topological zero modes anomalous, as dynamical space-time symmetries play a crucial role in enriching topological phases in Floquet systems.

\begin{figure}[t]
    \centering
    \includegraphics[width=\columnwidth]{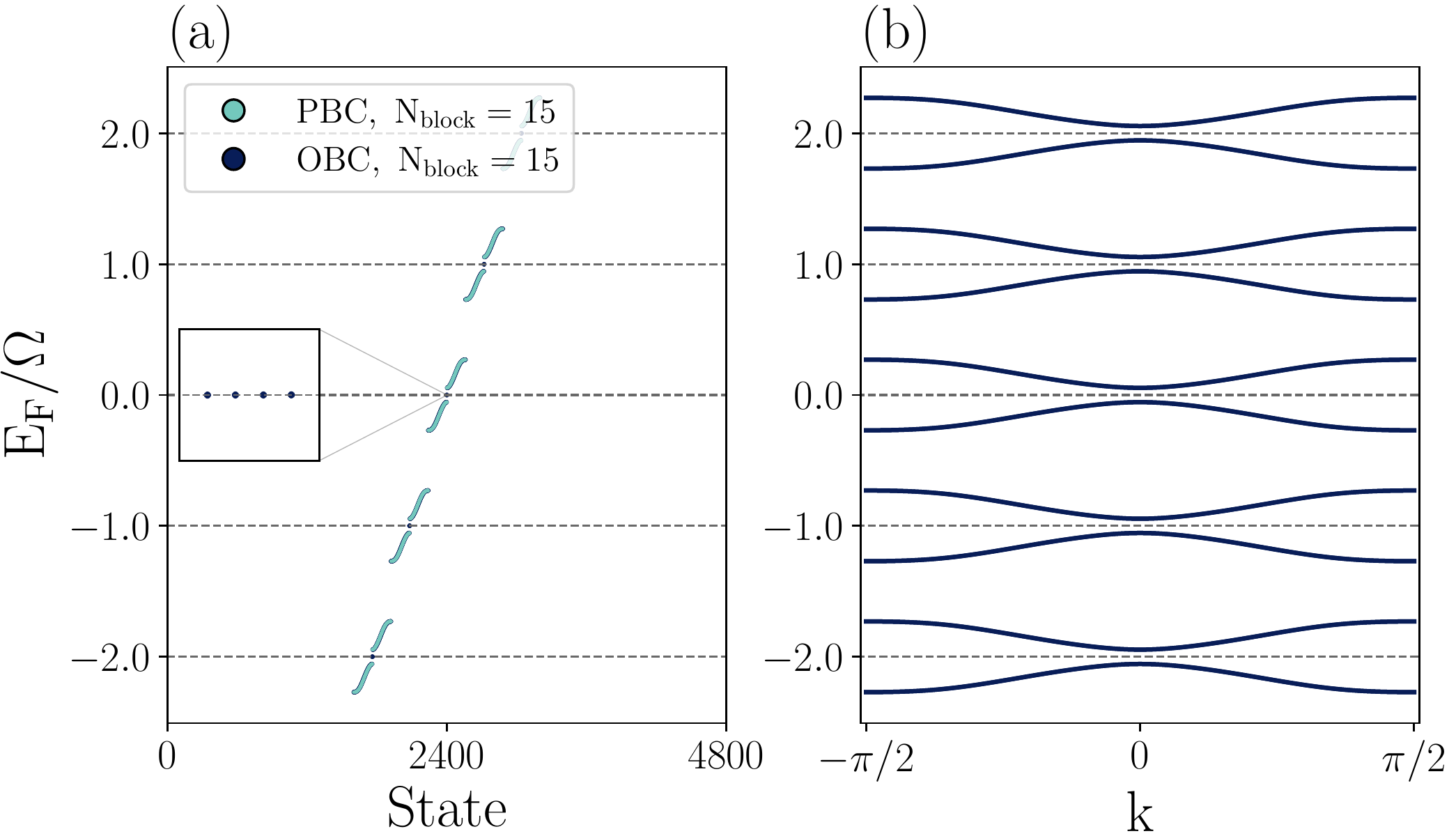}
    \caption{(a) The quasi-energy spectrum of the direct sum of two identical copies of $H(k,t)$ in Eq.~\eqref{eq:four-step driven model}. Specifically, we construct the truncated Floquet Hamiltonian $\mathcal{H}$ using the frequency-domain formulation in Eq.~\eqref{eq:Even/odd truncation of H_enlarged}. Inset: Anomalous topological zero edge modes appear within the zero gap, consistent with the expected topological invariant $\mathbb{Z}=2$. (b) The spectrum in periodic boundary conditions.  Here, the frequency indices $m,n$ run over a large, but finite range $-(N_{block}-1)/2\leq m,n\leq(N_{block}-1)/2$, where $N_{block}$ is much larger than the frequency space localization of the Floquet states. The parameters used are $T=1, J_1=\Omega=2\pi$, $J_2/J_1=0.9$, $J_3/J_1=J_4/J_1=0.1$, and $\rm N_{block}=15$. }
    \label{fig:Frequency_QES}
\end{figure} 

Notice that the topological invariant $\mathbb{Z}$ of class D in 1D with the reflection $\mathcal{U}$ anti-commuting with $\mathcal{C}$ which describes these zero modes, is the same as the invariant $M\mathbb{Z}_A$ of class A in 1D with the reflection. This is because $\mathcal{C}$ anti-commutes with the reflection $\mathcal{U}$, so that it exchanges the two different mirror subsectors, and thus each subsector does not have particle-hole symmetry. Focusing only on the subsector associated with reflection eigenvalue $R=+1$ (represented by a superscript $+$ below), the $\mathbb{Z}$ invariant can be expressed as
\begin{equation} \label{eq:MZ-invariant in 1D class D/A}
    \mathbb{Z}=M\mathbb{Z}^{+}_{A}=\nu^{+}_{k=0}-\nu^{+}_{k=\pm \pi/2}=-M\mathbb{Z}^{-}_{A},   
\end{equation}
where $\nu^{+}_{k^{*}}$ represents the number of negative energy modes of $\mathcal{H}(k)$ associated with reflection eigenvalue $R=+1$ at the zero-dimensional reflection invariant $k^{*}$ point.

We obtain $\mathbb{Z}=1$ from $H(k,t)$ in Eq.~\eqref{eq:four-step driven model}. In order to test this classification numerically, we take the direct sum of two identical copies of $H(k,t)$ and obtain four degenerate zero modes ($\mathbb{Z}=2$) in Fig.~\ref{fig:Frequency_QES}. Here, we realize topological zero modes by using parameters not in the high-frequency limit($\Omega\gg\left | J \right |)$, so we use large truncation blocks to ensure that the spectrum of the truncated Floquet Hamiltonian is a good approximation to the exact result within the first few quasi-energy zones centered around $E_F=0$~\cite{MarinBukov2015}.   

\emph{Frequency lattice decoration and symmetry group extensions}---The varying topological classifications at zero and $\pi$ gaps arise from the distinct classes of the extended symmetry group $\tilde{G}$, which fully describe the symmetry of the enlarged Hamiltonian $\mathcal{H}$ along the frequency lattice. The extended symmetry group $\tilde{G}$ is associated with each element of the second cohomology group $H^2[G,N]$, where $N$ is an Abelian group that defines the structure of decorated frequency lattice. For a given $G$, with subgroup $M$ and $A$ representing dynamical space-time symmetry and antisymmetry respectively, $\tilde{G}$ is the nontrivial extension of $G=M\times A$ (isomorphic to $\mathbb{Z}_2\times\mathbb{Z}_2$) by $N$. Two extensions can be considered equivalent if they correspond to the same element in the second cohomology group~\cite{SM}. 

The projective representation of symmetry and antisymmetry of the enlarged Hamiltonian $\mathcal{H}(k)$ are expressed as a unitary $\mathcal{U}$ and antiunitary $\mathcal{\bar{A}}$ in Eq.~\eqref{eq:Unitary,Antiunitary Syms in Enlarged Hamiltonian}. Each vertex of the frequency lattice is labeled with $n\in\mathbb{Z}$. The symmetry $\mathcal{U}$ is on-site, inducing a hidden color due to the alternating signs between even and odd sites. This hidden color is typically not observable because even and odd sites are interchangeable, but the antisymmetry $\mathcal{\bar{A}}$ can reveal the color on each vertex by its non-local mirror operation along the lattice, $\bar{A}_{0}h_n(k)\bar{A}^{-1}_{0}=-h_{-n}(k)$. The structure of the frequency lattice with a decoration represented by the colors on the lattice vertices is determined by an Abelian group $N\cong\mathbb{Z}\times H^1[A=\mathbb{Z}_2,M=\mathbb{Z}_2]=\mathbb{Z}\times\mathbb{Z}_{gcd(2,2)}$. This is generated by the $T$ time translations denoted by $U_F$ and colors denoted by $\bar{E}=e^{i2\pi/|H^1[A,M]|}=(-1)$~\cite{SM}:
\begin{equation} \label{eq:Identification of module N}
    N=\left \{ U_F^{a} \bar{E}^b|a\in \mathbb{Z}, b\in \mathbb{Z}_2\right \}.
\end{equation}

\begin{figure}[t]
    \centering
    \includegraphics[width=\columnwidth]{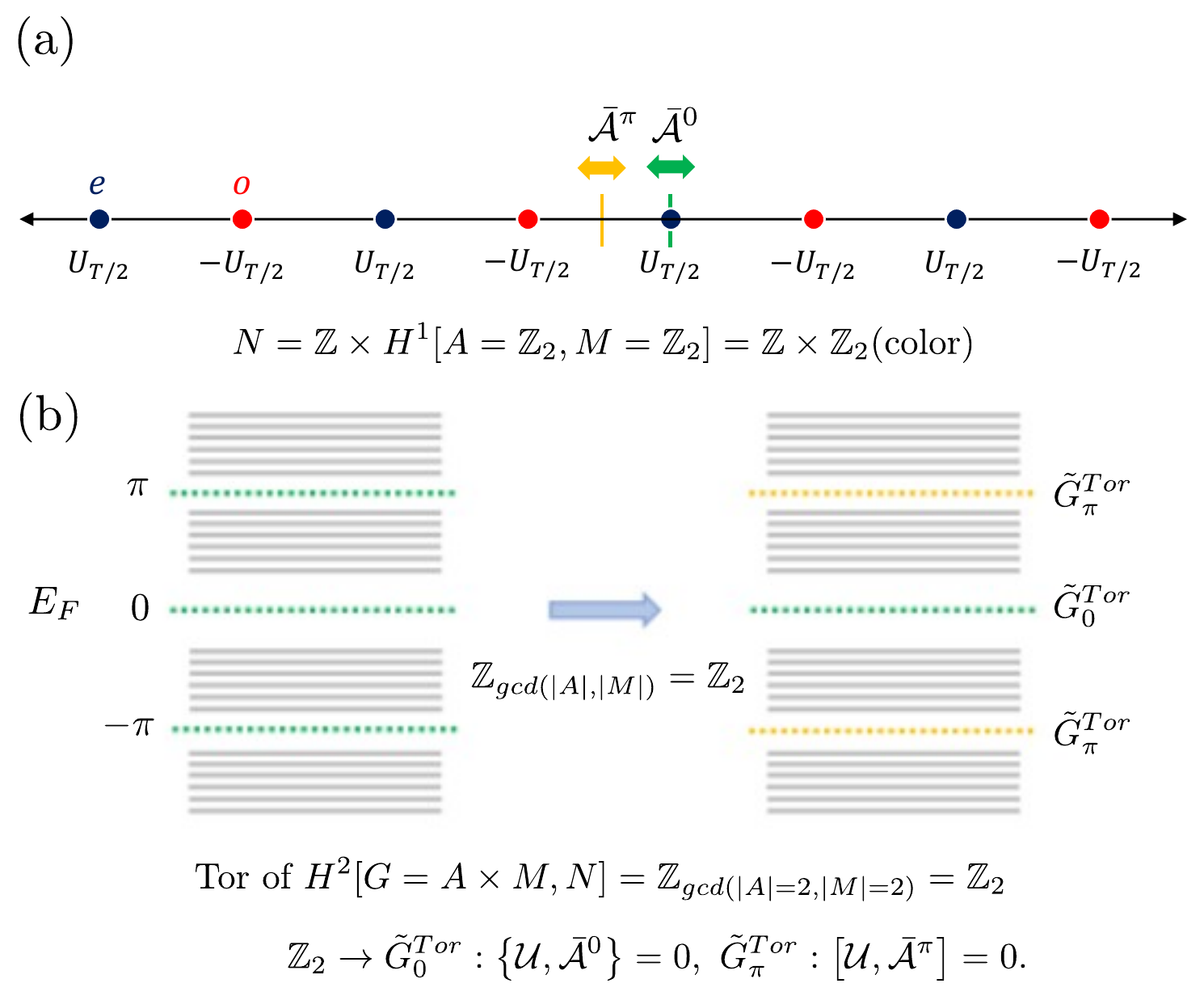}
    \caption{(a) The frequency lattice labels each site with $n\times\Omega$ for $n\in\mathbb{Z}$, representing the discrete $T$-time translation of $H(t)$. The interplay between the space-time symmetry $M$ and antisymmetry $A$ of the finite group $G=M\times A$ can be used to color the vertices of the lattice. The symmetry action of $M$ onto each vertex, but applied differently to even and odd sites alternatively(the order-two of symmetry $M$), which generates a hidden color. This color is not observable, but can be revealed by applying an additional non-local mirror operation of antisymmetry $A$ along the lattice. The lattice structure is determined by the Abelian group $N$ in Eq.~\eqref{eq:Identification of module N}, generated by the time-loop $U_F$ and colors $\bar{E}$, and is isomorphic to $\mathbb{Z}\times H^1[A,M]$. (b) Defining $\mathbb{Z}_2$ colors from $H^1[A=\mathbb{Z}_2,M=\mathbb{Z}_2]=\mathbb{Z}_2$, the torsion product of $H^2[G=\mathbb{Z}_2\times\mathbb{Z}_2,N]$ gives two different classes of group extensions, $\tilde{G}^{Tor}_0$ and $\tilde{G}^{Tor}_{\pi}$, distinguished by the different group algebra $[\mathcal{U},\mathcal{\bar{A}}^{0(\pi)}]_{+(-)}=0$, resulting in different topological classifications at the zero-gap and $\pi$-gap. }
    \label{fig:Frequency_lattice}
\end{figure}

\emph{Distinct topological classifications at different energy gaps}---The torsion product of the second cohomology group $H^2[G,N]$ provides a way to understand how a Floquet system can exhibit different topological classifications at different energy gaps in its quasi-energy spectrum. To generate these distinct classifications, the non-zero $\textrm{Tor}[A,M]$ (which is equivalent to $H^1[A,M]$ for finite Abelian groups $M$ and $A$) is important. To observe different topological classifications, it is necessary to assign colors (nontrivial $\bar{E}\neq1$) to the lattice vertices on the frequency lattice. We notice that the torsion product of $H^2[G=\mathbb{Z}_2\times\mathbb{Z}_2,N=\mathbb{Z}\times\mathbb{Z}_2]$ is given by $\textrm{Tor}[H^3(G,\mathbb{Z}),N]=\mathbb{Z}_2$. This represents the commutation and anti-commutation relations between two projective representations $\mathcal{U}$ and $\mathcal{\bar{A}}$ denoted by $c=\pm 1$ and can be written as
\begin{equation} \label{eq:torsion product of the group extension}
    \mathcal{U} \mathcal{\bar{A}}~\mathcal{U}^{-1}\mathcal{\bar{A}}^{-1}=\bar{E}^{c}.
\end{equation}
Notice that the antisymmetry $\mathcal{\bar{A}}$ acts on the frequency lattice as a mirror with two possible centers: either passing through the lattice vertex or through the middle of the bond between two vertices, denoted as $\mathcal{\bar{A}}^0$ or $\mathcal{\bar{A}}^{\pi}$, respectively. Combined with the space-time symmetry $\mathcal{U}$, the two types of mirror operations $\mathcal{\bar{A}}^{0,\pi}$ can be distinguished such that $\mathcal{\bar{A}}^{\pi}$ exchanges the colors $(e\leftrightarrow o)$, whereas $\mathcal{\bar{A}}^0$ does not. This results in two different classes of group extensions, $\tilde{G}^{Tor}_0$ and $\tilde{G}^{Tor}_{\pi}$ as shown in Fig.~\ref{fig:Frequency_lattice}, depending on how $\mathcal{U}$ and $\mathcal{\bar{A}}$ interact. These classes are distinguished by the different group algebras $[\mathcal{U},\mathcal{\bar{A}}^{0(\pi)}]_{+(-)}=0$ which arise from the opposite sign of $\bar{E}^c$ in Eq.~\eqref{eq:torsion product of the group extension}. This leads to different topological classification at zero and $\pi$ gaps as $\mathbb{Z}(\mathbb{Z}_1)$ in our model. 

More generally, for the 1D frequency lattice with only order-two antisymmetry $\mathcal{\bar{A}}$ in Floquet systems, at most two distinct topological classifications can occur at the zero-gap and $\pi$-gap, when there coexists an even order-$D$ space-time symmetry. This follows from the torsion product of $H^2[G,N]$, which is given by~\cite{Mesajos2013,SM}   
\begin{align} \label{Torsion calculations}
    &\textrm{Tor}\left[H^2[G=\mathbb{Z}_D\times \mathbb{Z}_2,U(1)],N=\mathbb{Z}\times H^1[\mathbb{Z}_2,\mathbb{Z}_{D}]\right] \nonumber \\
    &=\textrm{Tor}[\mathbb{Z}_{gcd(2,D)},\mathbb{Z}_{gcd(2,D)}]=\mathbb{Z}_{gcd(2,D)},  
\end{align}
where $N$ is identified as $\mathbb{Z}\times\mathbb{Z}_{gcd(2,D)}$ from $H^1[A,M]=H^1[\mathbb{Z}_2,\mathbb{Z}_D]=\textrm{Tor}[\mathbb{Z}_2,\mathbb{Z}_{D}]=\mathbb{Z}_{gcd(2,{D})}$. This indicates that different topological classifications cannot occur in a Floquet system with odd-order space-time symmetry.

\emph{Conclusion and Outlook}---In this work, we demonstrate that when considering order-two space-time symmetries and antisymmetries, distinct topological classifications can occur at different energy gaps in the Floquet quasi-energy spectrum. We present a simple 1D model that preserves both a time-glide symmetry and an antiunitary antisymmetry.
Using the frequency-enlarged Hamiltonian, we found that the model exhibits anomalous $\mathbb{Z}$-classified topological zero modes, and is trivial at $\pi$-gaps. The origin of this difference lies in the $T/2$ half-translation component of the time-glide action, which results in opposite commutation relations between the effective reflection and antisymmetry in the enlarged Hamiltonian. We confirm this difference by showing that the torsion product of the second cohomology group $H^2[G,N]$ accounts for the distinct topological classification, which is further used to predict that such an energy-dependent classification can also exist if we replace the order-two space-time symmetry by an even-order one. 

Two generalizations of the current work are worth mentioning to explore in the future.
It is known that the 1D frequency lattice perspective of the Floquet system can be generalized to quasi-periodically driven systems, where a higher-dimensional frequency lattice representation is obtained~\cite{Martin2017,Yang2018, Peng2018, Crowley2019}.
The effects of the antisymmetry on this frequency lattice will be more complicated, and may allow for richer and more distinct topological classifications. Moreover, we leave the question of how topological classifications~\cite{Potter2016,Else2016} vary at different energy gaps in the presence of interactions and space-time symmetries/antisymmetries, as well as how to realize those stably~\cite{Potirniche2017,Else2017}, for future work.  

\emph{Acknowledgement.}---I. N. and J. K. thank Philip Crowley, Sin\'{e}ad M. Griffin, Francisco Machado, Kazuaki Takasan for stimulating discussions on related projects. I. N. is supported by the US Department of Energy, Office of Science, National Quantum Information Science Research Centers, Quantum Systems Accelerator (QSA). J. K. acknowledges support from the Air Force Office of Scientific Research via the MURI program (FA9550-21-1-0069).  R. ~J.~ S acknowledges funding from a New Investigator Award, EPSRC grant EP/W00187X/1, as well as Trinity College, Cambridge.
 Y.P. is supported by the NSF PREP grant (PHY-2216774).

%\bibliography{manuscript}

\clearpage
\onecolumngrid
\section*{Supplementary material for: Distinct Floquet topological classifications from color-decorated frequency lattices with space-time symmetries}
\twocolumngrid
\section{Symmetrized unitary loop operators}
For a given Hamiltonian $H(t)$ with the Floquet period $T$, the time-evolution operator $U(t)=U(t+t_0,t_0)=T~\exp\left [-i \int_{t_0}^{t_0+t} dt^{\prime} H(t^{\prime})\right ]$ by setting its initial time $t_0=0$, can be decomposed as $U(t)=V(t)e^{-iH_Ft}$ where $H_F=\sum_{n=1}^{N} (E_F)_n \ket{\Phi_n}\bra{\Phi_n}$. The set \{$(E_F)_n$\} is referred to as the Floquet quasienergy spectrum within $\left [-\pi/T,\pi/T\right ]$ by setting the branch cut at $-\pi/T$ without loss of generality. We define $V(t)=\sum_n \ket{\Phi_n(t)}\bra{\Phi_n(0)}$ as the unitary loop operator satisfying $\ket{\Phi_n(t+T)}=\ket{\Phi_n(t)}$ and  \{$\ket{\Phi_n(t=0)}$\} is the set of Floquet eigenstates \{$\ket{\Phi_n}$\}. $U(t)$ may be constructed differently as
\begin{equation} \label{eq:symmetrized t-evolution operator}
    U_{\tau}(t)=T~\exp[-i\int_{\tau-t/2}^{\tau+t/2} dt^{\prime} H(t^{\prime})],
\end{equation}
upon centering around time $\tau$. It has the same quasi-energy spectrum regardless of the choice of $\tau$ from $U_{\tau}(T)=SU_0(T)S^{\dag}$, where $S=U(\tau+T/2)U^{\dag}(T/2)$ is the unitary operator. $U_{\tau}(t)$ gives rise to the same topological classification as $U(t)$ as these are in principle gauge related. In Ref.~\cite{Peng2020}, the author classified the unitary loop operator $V_\tau(t)$, $V_\tau(t=n \cdot T)=\mathbb{I}$ for $n\in\mathbb{Z}$, whose singularity could give rise to the anomalous Floquet topological phases~\cite{Rudner2013}, by setting it equal to $U_{\tau}(t)$ in Eq.~\eqref{eq:symmetrized t-evolution operator}, from the assumption of a spectral-flattened Floquet Hamiltonian $H_F=0$.

We construct the symmetrized unitary loop $V_{\tau}(t)$ as
\begin{equation} \label{eq:symmetrized loop operator}
    V_{\tau}(t)=\sum_{n=1}^{N}\ket{\Phi_n(\tau+t/2)}\bra{\Phi_n(\tau-t/2)},
\end{equation}
without the assumption of flattening ($H_F=0$) such that the above can be applied to any given Floquet systems. This operator satisfies $V_{\tau}(t=n\cdot T)=\mathbb{I}$ for any integer n $\in\mathbb{Z}$ with $V_{\tau}(t)=V_{\tau}(t+2T)$ and $V_{\tau}(t)=V_{\tau+T/2}^{\dag}(T-t)$ from the loop characteristic of $V_{\tau}(t)$. For noninteracting Floquet systems with well-defined momentum $\mathit{\mathbf{k}}$, the matrix representation of $V_{\tau}(k,t)$ can be written as $V_{\tau}(\mathit{\mathbf{k}},t)=\sum_{\alpha,\beta=1}^{N}\left[V_{\tau}(\mathit{\mathbf{k}},t)\right ]_{\alpha,\beta}~c^{\dag}_{\mathit{\mathbf{k}},\alpha}c_{\mathit{\mathbf{k}},\beta}$ with respect to internal degrees of freedom $\alpha,\beta$ of the free-fermion constituents.

\section{Symmetries in Floquet systems}

The action of symmetry operations of time-reversal $T$, particle-hole $C$, chiral $S$ symmetries represented as $U_TK$, $U_CK$, and $U_S$ respectively, on the symmetrized unitary loop $V_{\tau}(\mathit{\mathbf{k}},t)$ can be summarized as
\begin{align}
    U_TV_{\tau}(\mathit{\mathbf{k}},t)^{*}U_T^{\dag} =& V_{-\tau}^{\dag}(-\mathit{\mathbf{k}},t), \label{eq:local symmetry representation} \\
    U_CV_{\tau}(\mathit{\mathbf{k}},t)^{*}U_C^{\dag} =& V_{\tau}(-\mathit{\mathbf{k}},t), \nonumber \\
    U_SV_{\tau}(\mathit{\mathbf{k}},t)U_S^{\dag} =& V_{-\tau}^{\dag}(\mathit{\mathbf{k}},t), \nonumber
\end{align}
where $T^2=U_TU_T^{*}=\epsilon_T=\pm\mathbb{I}$, $ C^2=U_CU_C^{*}=\epsilon_C=\pm\mathbb{I}$, and $S^2=U_S^2=\epsilon_S=\mathbb{I}$. In addition to the non-spatial symmetries, the action of space-time unitary symmetry/antiunitary antisymmetry operators $O$ and $\bar{O}$, on the Hamiltonian and the symmetrized unitary loop $V_{\tau}(\mathit{\mathbf{k}},t)$ can be summarized as
\begin{align}
    U_sH(\mathit{\mathbf{k}},t)U_s^{\dag} =& H(-\mathit{\mathbf{k_{\parallel}}},\mathit{\mathbf{k_{\perp}}},t+s), \label{eq:non-local space-time symmetry representation} \\
    \bar{A}_sH(\mathit{\mathbf{k}},t)\bar{A}_s^{-1} =& -H(\mathit{\mathbf{k_{\parallel}}},-\mathit{\mathbf{k_{\perp}}},-t+s), \nonumber \\
    U_sV_{\tau}(\mathit{\mathbf{k}},t)U_s^{\dag} =& V_{\tau+s}(-\mathit{\mathbf{k_{\parallel}}},\mathit{\mathbf{k_{\perp}}},t), \nonumber \\
    \bar{A}_sV_{\tau}(\mathit{\mathbf{k}},t)\bar{A}_s^{-1} =& V_{\tau+s}(\mathit{\mathbf{k_{\parallel}}},-\mathit{\mathbf{k_{\perp}}},t), \nonumber
\end{align}
where $\mathit{\mathbf{k_{\parallel}}}$ and $\mathit{\mathbf{k_{\perp}}}$ denote the components of the momentum which flip under the symmetry/antisymmetry respectively. The $s$ can take on values of either $0$ or $T/2$ for $U_s$ and $\bar{A}_s$ owing to the periodicity in $t$ and the order-two nature of the space-time symmetry $O$/antisymmetry $\bar{O}$. We set $\tau=T/2$ without loss of generality and omit the subscript $\tau$ from $V_{\tau}(\mathit{\mathbf{k}},t)$ from now on. Consequently, we have the relation $U_{s=T/2}V(\mathit{\mathbf{k}},t)U^{\dag}_{s=T/2}=V^{\dag}(-\mathit{\mathbf{k_{\parallel}}},\mathit{\mathbf{k_{\perp}}},T-t)$, which arises from $V_{0}(t)=V^{\dag}_{T/2}(T-t)$. Furthermore, we obtain $\bar{A}_{s=0}V(\mathit{\mathbf{k}},t)\bar{A}^{-1}_{s=0}=V(\mathit{\mathbf{k_{\parallel}}},-\mathit{\mathbf{k_{\perp}}},t)$.

\section{Classification of Loop operators with Hermitian Map}

The effective Hamiltonian $H_{\text{eff}}(\mathbf{k},t)$ can be constructed using the symmetrized unitary loop $V(t)$ through the Hermitian map~\cite{Roy2017,Morimoto2017}. The specific form of the Hamiltonian depends on whether $V(\mathbf{k},t)$ possesses chiral symmetry or not. In the case where $V(\mathbf{k},t)$ has chiral symmetry, the effective Hamiltonian is given by
\begin{equation} \label{eq:hermitian map1}
    H_{\text{eff}}(\mathit{\mathbf{k}},t)=U_SV(\mathit{\mathbf{k}},t),
\end{equation}
where the Hermitian map eliminates the existing chiral symmetry in $V(\mathbf{k},t)$ from $H_{\text{eff}}(\mathbf{k},t)$. For non-chiral symmetric loop $V(\mathit{\mathbf{k}},t)$, the effective Hamiltonian is given by
\begin{equation} \label{eq:hermitian map2}
    H_{\text{eff}}(\mathit{\mathbf{k}},t)=\begin{pmatrix}
                               0  &  V(\mathit{\mathbf{k}},t)\\
                               V^{\dag}(\mathit{\mathbf{k}},t) & 0 \\
                             \end{pmatrix}.
\end{equation}
where $H_{\text{eff}}(\mathit{\mathbf{k}},t)$ acquires a new effective chiral symmetry $\tilde{U}_SH_{\text{eff}}(\mathit{\mathbf{k}},t)\tilde{U}^{\dag}_S=-H_{\text{eff}}(\mathit{\mathbf{k}},t)$. Here $\tilde{U}_S=\rho_z \otimes \mathbb{I}$ with $\rho_{x,y,z}$ represent the Pauli matrices in the enlarged space.

\begin{figure}[t]
    \centering
    \includegraphics[width=\columnwidth]{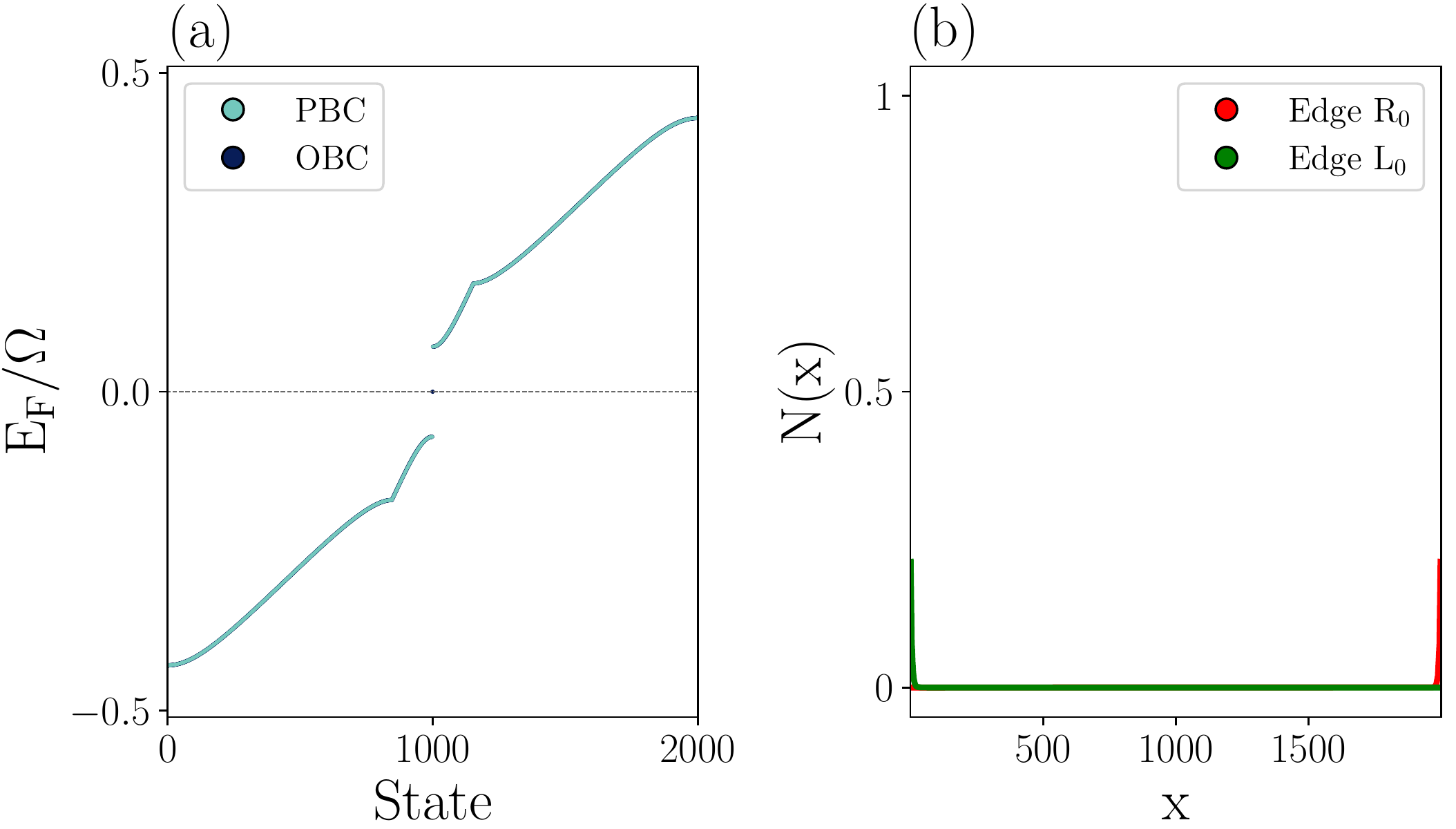}
    \caption{(a) Quasienergy spectrum of the 1D model with order-two antiunitary antisymmetry and time glide symmetry. (b) Localized topological edge states that appear within the zero gap. We use the parameters $J_1=\Omega=2\pi$(by setting $T=1$), $J_2/J_1=0.9$, $J_3/J_1=J_4/J_1=0.1$, and $L=2000$. Note that quasienergies in the "first Floquet zone" are $E_F/\Omega\in[-0.5,0.5]$.}
    \label{fig:Edge}
\end{figure}

We introduce a model of spinless fermions hopping on a bipartite one-dimensional lattice, which hosts anomalous Floquet topological zero modes protected by both a time-glide $U_{T/2}$ and an antisymmetry $\bar{A}_{0}$. The Hamiltonian is a four-step drive where $H_1(k)=J_1\sigma_x+[\delta-2J_4\sin(2k)]\sigma_z$ and $H_2(k)=-J_2\sin(2k)\sigma_x-J_2\cos(2k)\sigma_y-2J_3\sin(2k)\sigma_z$. Here, the Pauli matrices $\sigma_i$ act on the sublattice degrees of freedom. This Hamiltonian satisfies $U_{T/2}H_{1,2}(k)U^{\dag}_{T/2}=H_{3,4}(-k)$, $U_{T/2}=\sigma_x$ and $\bar{A}_{0}H(k,t)\bar{A}^{-1}_{0}=-H(k,t)$, $\bar{A}_0=i\sigma_yK$ where $K$ is the complex conjugation. Here, $U_{T/2}^2=\mathbb{I}$, $\bar{A}_0^2=-\mathbb{I}$, and $\left \{U_{T/2},\bar{A}_0\right \}=0$. Its quasi-energy spectrum shows a bulk gap around $E_F=0$, where nontrivial edge states appear in Fig.~\ref{fig:Edge}.

The topological classification of the symmetrized unitary loop $\hat{V}(t)$ is equivalent to that of $H_{\textrm{eff}}(k,t)$ in Eq.~\eqref{eq:hermitian map2}. The Hermitian mapped Hamiltonian $H_{\textrm{eff}}(k,t)$ belongs to class AIII in 2D, where $t$ is treated as an additional spatial dimension. In addition to the chiral symmetry $\Tilde{U}_S$, $H_{\textrm{eff}}(k,t)$ has two additional order-two crystalline symmetries, given by
\begin{align}
    &\tilde{U}^{+}_{-}H_{\textrm{eff}}(k,t)(\tilde{U}^{+}_{-})^{\dag}=H_{\textrm{eff}}(-k,T-t)  \label{eq:Order-2 symmetries in the Hermitian mapped H}, \\
    &\tilde{A}^{-}_{+}H_{\textrm{eff}}(k,t)(\tilde{A}^{-}_{+})^{-1}=H_{\textrm{eff}}(k,t), \nonumber
\end{align}
where $\tilde{U}^{+}_{-}=\rho_x\otimes U_{T/2}$ and $\tilde{A}^{-}_{+}=\rho_0 \otimes \bar{A}_0$ satisfy $\left \{ \tilde{U}_S,\tilde{U}^{+}_{-} \right \}=\left [\tilde{U}_S,\tilde{A}^{-}_{+}  \right ]=0$. Here, $(\tilde{U}^{+}_{-})^2=\mathbb{I}$, $(\tilde{A}^{-}_{+})^2=-\mathbb{I}$, and $\left \{ \tilde{U}^{+}_{-},\tilde{A}^{-}_{+} \right \}=0$. The classification of $H_{\textrm{eff}}(k,t)$, given by $K_{R}^{U}(7,1;0,0,0,0)=\pi_0(R_6)=\mathbb{Z}_1 \equiv 0$~\cite{Shiozaki2014}, does not give rise to nontrivial topological phases in terms of $\pi$ modes, which is in contrast to the $\mathbb{Z}$ classification as obtained from the more standard frequency-domain formulation of topological zero modes.

\section{1D Class A with time-glide symmetry}

We evaluate the topological classification of a 1D class A Hamiltonian with solely time-glide symmetry by examining its Hermitian mapped effective Hamiltonian $H_{\textrm{eff}}(\mathit{\mathbf{k}},t)$ in Eq.~\eqref{eq:hermitian map2}. Here, the $H_{\textrm{eff}}(k,t)$ describes a static 2D class AIII with an additional spatial inversion symmetry $\tilde{U}^{+}_{-}$ given by $\tilde{U}^{+}_{-}H_{\textrm{eff}}(k,t)(\tilde{U}^{+}_{-})^{\dag}=H_{\textrm{eff}}(-k,T-t)$, where $\tilde{U}^{+}_{-}=\rho_x\otimes U_{T/2}$ and $(\tilde{U}^{+}_{-})^2=\mathbb{I}$. The $\tilde{U}^{+}_{-}$ anti-commutes with the chiral symmetry $\tilde{U}_S=\rho_z\otimes\mathbb{I}$, allowing for a $\mathbb{Z}$ classification~\cite{Shiozaki2014}.

\begin{figure}[t]
    \centering
    \includegraphics[width=\columnwidth]{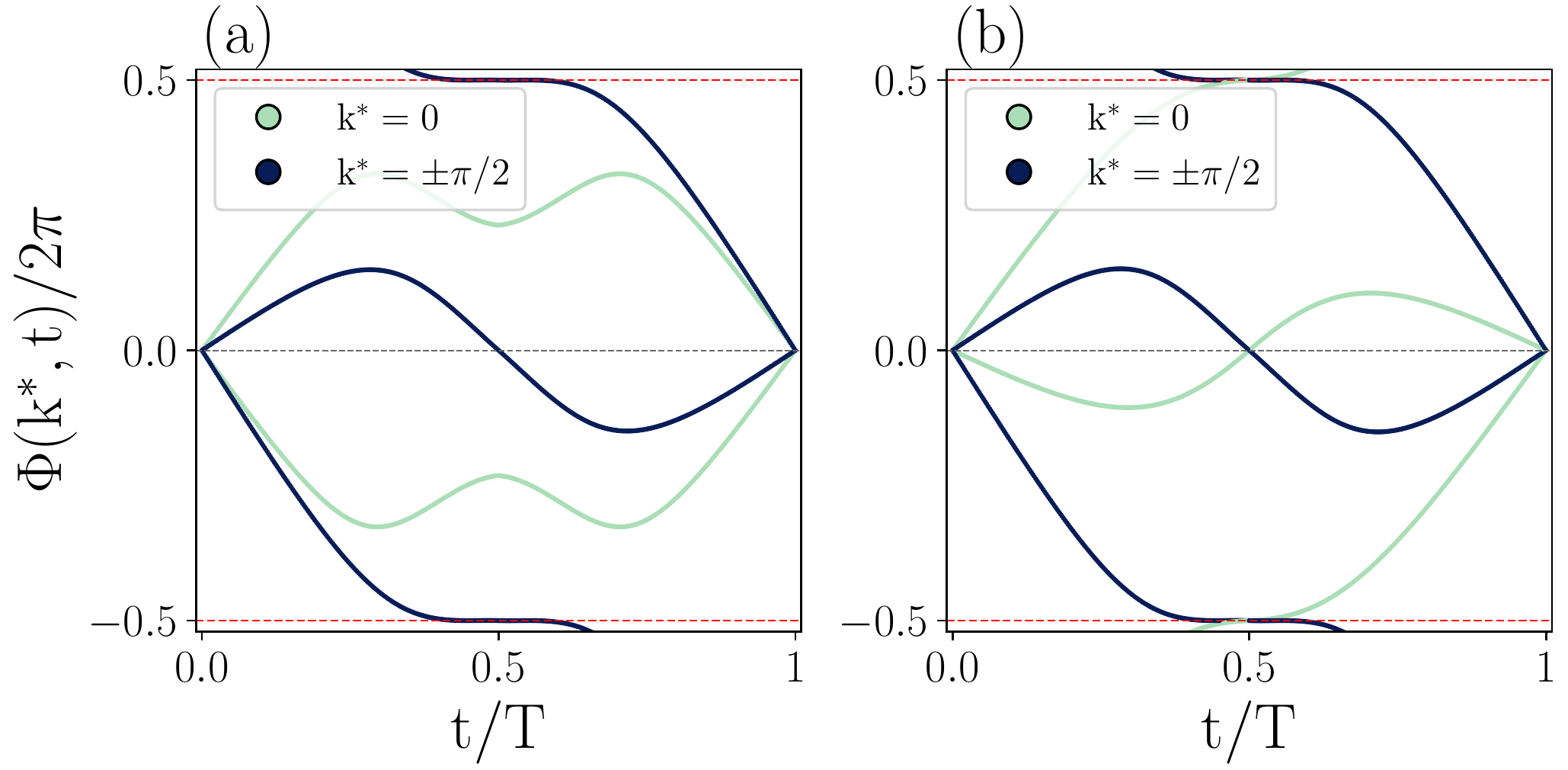}
    \caption{(a) Phase band $\phi_{n=1,2}(k,t)$ of the symmetrized unitary loop $V(k,t)$ in Eq.~\eqref{eq:symmetrized loop operator} at $k^{*}=0,\pm \pi/2$ for $\mathbb{Z}_{loop}=1, \nu_{k^{*}=0}=0, \nu_{k^{*}=\pm \pi/2}=-1$ at $J_2/J_1=2.4$ and (b) $\mathbb{Z}_{loop}=2, \nu_{k^{*}=0}=1, \nu_{k^{*}=\pm \pi/2}=-1$ at $J_2/J_1=2.1$. We use the parameters $J_1=\Omega=2\pi$($T=1$), $J_3/J_1=J_4/J_1=0.1$ of the modified $H(k,t)$ with only the time-glide symmetry.}
    \label{fig:phase_band}
\end{figure}

The $\mathbb{Z}_{loop}$ invariant, which arises from the singularity of the symmetrized loop $V(t)$, can be obtained as $\mathbb{Z}_{loop}=\nu_{k=0}-\nu_{k=\pm\pi/2}=\nu_{t=0}-\nu_{t=T/2}$ where $\nu_{k^{*}}$ and $\nu_{t^{*}}$ are the winding number along $t$ and $k$ given by
\begin{align} \label{eq:Z-invariant from the loop}
    \nu_{k^{*}}&=\frac{i}{T}\int_{0}^{T}dt\Tr{\left [ V^{\dag}(k^{*},t)\partial_tV(k^{*},t)\right ]}, \nonumber \\
    \nu_{t^{*}}&=\frac{i}{2\pi}\int_{-\pi/2}^{\pi/2}dk\Tr{\left [ V^{\dag}(k,t^{*})\partial_kV(k,t^{*})\right ]}.
\end{align}
From the $V(k,t=n\cdot T)=1$ in Eq.~\eqref{eq:symmetrized loop operator}, the $\nu_{t^{*}=0}$ becomes zero. The $\mathbb{Z}_{loop}\mod2$ can be interpreted as the Kane-Mele $Z_2$ index $(-1)^{\nu}$~\cite{Fu2007}, given that $(-1)^{\mathbb{Z}_{loop}}=(-1)^{\nu}$. To ensure the existence of only the time-glide symmetry $U_{T/2}$, we modify the original four-step driven model preserving both $U_{T/2}$ and $\bar{A}_0$. The modified Hamiltonians are given by $H_1(k)=2J_3\cos(2k)\sigma_0+J_1\sigma_x+[\delta-2J_4\sin(2k)]\sigma_z, H_2(k)=2J_3\cos(2k)\sigma_0-J_2\sin(2k)\sigma_x-J_2\cos(2k)\sigma_y$. The evaluated $\mathbb{Z}_{loop}=1$ in Eq.~\eqref{eq:Z-invariant from the loop} corresponds to the number of `zone-edge' singularities~\cite{Nathan2015} in the phase band $\left \{ \phi_n(k,t) \right \}$ of symmetrized unitary loop $V(k,t)=\sum_n e^{-i\phi_n(k,t)}\ket{n(k,t)}\bra{n(k,t)}$, which are protected by the above-defined time glide symmetry imposing $\phi_n(k,t)=-\phi_n(-k,T-t)\mod 2\pi$ (see Fig.~\ref{fig:phase_band}).

The same classification result can be obtained from the frequency-domain formulation by classifying the static frequency enlarged Hamiltonian $\mathcal{H}(k)$. This Hamiltonian resides in class A of the ten-fold way, but also has an additional reflection symmetry $\mathcal{U}$, which arises from the time-glide symmetry $U_{T/2}$ of $H(k,t)$. The frequency-domain formulation allows for the existence of $\mathbb{Z}$-classified topological edge states that can appear at any gap in the quasi-energy spectrum of the truncated $\mathcal{H}$, regardless of the even/odd block dimension. The $\mathbb{Z}$ invariant characterizing the topological edges at the quasi-energy gap $E_F=0$ can be evaluated as $M\mathbb{Z}^{+}_{A}=\nu^{+}_{k=0}-\nu^{+}_{k=\pm \pi/2}$ where $\nu^{+}_{k^{*}}$ represents the number of negative energy modes of $\mathcal{H}(k)$ associated with the reflection eigenvalue $R=+1$ at zero-dimensional reflection invariant $k^{*}$ point. We obtain the $M\mathbb{Z}^{+}_{A}=1$, which is consistent with the $\mathbb{Z}_{loop}=1$ obtained from Eq.~\eqref{eq:Z-invariant from the loop}.

\section{Frequency lattice decoration and symmetry group extensions}
In the main text, we discuss the classification of the extended space-time group $\tilde{G}$ of the frequency-enlarged Hamiltonian $\mathcal{H}$, which arises from the symmetry group $G$ of $H(t)$ and is associated with each element of the second cohomology group $H^2[G,N]$, where $N$ is an Abelian group that defines the structure of decorated frequency lattice (c.f. also ~\cite{Chen2014} for slightly related equilibrium pursuits). More specifically, the extended symmetry group $\tilde{G}$ can be viewed as a nontrivial extension of $G$ by $N$, and it is described by the following exact sequence:
\begin{equation} \label{eq:group_extension_exact_sequence}
1\rightarrow N\rightarrow \tilde{G}\rightarrow G\rightarrow 1.
\end{equation}
Two extensions can be considered equivalent if they correspond to the same element in the second cohomology group. We provide a general demonstration of the connection between the equivalence classes of group extensions $\tilde{G}$ and the second cohomology group~\cite{Xie_chen2013,Alexandradinata2016}.

\subsection{Connection between group extensions and the second cohomology group}
For a group $G$, let $N$ be a $G$-module, which is an Abelian group on which $G$ acts compatibly with the multiplication operation on $N$ as $g\cdot(ab)=(g\cdot a)(g\cdot b)$ for every $g\in G$ and $a,b\in N$. Letting the group element $g\in G$ be represented in $\tilde{G}$ by $\tilde{U}_g$, the automorphism induced by $g$ as $a\rightarrow(g\cdot a)=\sigma_{g}(a)\in N$(normal subgroup of $\tilde{G}$) can be written as $\sigma_g(a)=\tilde{U}_ga\tilde{U}_g^{-1}$ satisfying
\begin{equation} \label{eq:automorphism induced by G}
    \sigma_{g}(ab)=\sigma_{g}(a)\sigma_{g}(b),
\end{equation}
for every $a,b\in N$. The $\tilde{U}(g)$ forms a projective representation of $G$ with a corresponding factor $w(g_i,g_j)\in N$ for any given $g_i,g_j\in G$ as
\begin{equation} \label{eq:factor of projective rep}
    \tilde{U}_{g_i}\tilde{U}_{g_j}=w(g_i,g_j)\tilde{U}_{g_{ij}=g_ig_j},
\end{equation}
where $w(g_i,g_j)=\nu_2(\mathbb{I},g_i,g_{ij})$ corresponds to a 2-cochain with its first argument in $\nu_2\in C^2[G,N]$ as the identity element in $G$. If $w(g_i,g_j)=1$ for $\forall g_i,g_j\in G$, the projective representation $\tilde{U}_g$ reduces to the usual linear representation of $G$. Each factor transforms under $G$ as $\nu_2(gg_0,gg_1,gg_2)=\tilde{U}_g\nu_2(g_0,g_1,g_2)\tilde{U}_g^{-1}=\sigma_{g}(\nu_2(g_0,g_1,g_2))\in N$. The associativity with respect to the multiplication of the $\tilde{U}_g$'s, induces the factor to be as a 2-cocycle $\nu_2(g_0,g_1,g_2)\in\mathcal{Z}^2[G,N]=\left \{\nu_2|d_2\nu_2=\mathbb{I},\nu_2\in C^2[G,N]\right \}$~\cite{Xie_chen2013}.

Two projective representations $\tilde{U}_g$ and $\tilde{U}^{\prime}_g$ related by a gauge transformation $\tilde{U}_g\rightarrow\tilde{U}^{\prime}_g=\nu_1(\mathbb{I},g)^{-1}\tilde{U}_g$ with $\nu_1(\mathbb{I},g)^{-1}\in N$ and $\tilde{U}^{\prime}_{g_i}\tilde{U}^{\prime}_{g_j}=\nu_2^{\prime}(\mathbb{I},g_i,g_{ij})\tilde{U}^{\prime}_{g_{ij}}$, are gauge-equivalent when they induce the same automorphism on $N$ such that $\sigma_{g}(a)=\tilde{U}_g a \tilde{U}^{-1}_g=\tilde{U}^{\prime}_g a {\tilde{U^{\prime}}}^{-1}_g$. Thus, two equivalent 2-cochains $\nu_2(g_0,g_1,g_2)$ and $\nu_2^{\prime}(g_0,g_1,g_2)$ differ only by multiplication with a 2-coboundary $d_1\nu_1(g_0,g_1,g_2)\in \mathcal{B}^2[G,N]=\left\{\nu_2|\nu_2=d_1\nu_1|\nu_1\in C^1[G,N]\right \}$ as~\cite{Xie_chen2013,Alexandradinata2016},
\begin{align} \label{eq:gauge-equivalent two cochains}
    \nu_2(g_0,g_1,g_2)&=\nu_2^{\prime}(g_0,g_1,g_2)\nu_1(g_1,g_2)\nu_1(g_0,g_1)\nu_1(g_0,g_2)^{-1} \nonumber \\
    &=\nu_2^{\prime}(g_0,g_1,g_2)[d_1\nu_1](g_0,g_1,g_2).
\end{align}
Therefore, the different equivalence classes of projective representations are determined by the elements of the second cohomology group $H^2[G,N]=\mathcal{Z}^2[G,N]/\mathcal{B}^2[G,N]$, which correspond to the equivalence classes of 2-cocycles modulo 2-coboundaries. The identity element $\mathbb{I}\in H^2[G,N]$ represents the class that corresponds to the linear representation of the group $G$, and its extension $\tilde{G}$ becomes the semi-direct product of $N$ and $G$, written as $\tilde{G}=N\rtimes G$.

\subsection{Identification of N and space-time group extensions}
We now demonstrate how the group $H^2[G,N]$ with coefficients in $N$ characterizes the equivalence classes of the extended symmetry group $\tilde{G}$ that fully describe the symmetry of the enlarged Hamiltonian $\mathcal{H}$ along the frequency lattice. The lattice has each site labeled with $n\cdot\Omega$ with $n\in\mathbb{Z}$, representing the discrete $T$-time translation of $H(t)$. Furthermore, vertices of the lattice can be colored based on the interplay between the space-time symmetry $M$ and antisymmetry $A$ of the given finite group $G=M\times A$. While $A$ acts as a non-local mirror operation along the frequency lattice, the symmetry action of $M$ is on-site, but alternates in sign between even and odd sites (when we consider the order-two space-time symmetry $M$), which generate a hidden color. This hidden color is typically not observable because even and odd sites are interchangeable, but the antisymmetry $A$ can reveal the color on each vertex. The structure of the frequency lattice is determined by an Abelian group $N\cong\mathbb{Z}\times H^1[A,M]$. This is generated by the $T$ time translations denoted by $U_F$ and colors denoted by $\bar{E}=e^{i2\pi/|H^1[A,M]|}$:
\begin{equation} \label{eq:Identification of module N}
    N=\left \{ U_F^{a} \bar{E}^b|a\in \mathbb{Z}, b\in \mathbb{Z}_{|H^1[A,M]|}\right \}.
\end{equation}
The automorphism induced by $g\in G$ on $n=U_F^a\bar{E}^b\in N$ is denoted as $(g\cdot n)$ and can be written as
\begin{align} \label{eq:automorphism by G on N}
    \sigma_g(U_F^{a} \bar{E}^b)&=\tilde{U}_g(U_F^{a} \bar{E}^b)\tilde{U}_g^{-1} \nonumber \\
    &=U_F^{\kappa(g)\cdot a}\bar{E}^b~~\rm with~\kappa(g)\in\left \{\pm 1\right \},
\end{align}
where $\kappa(g)=1$ preserves the orientation of the time-loop when $\tilde{U}_g$ corresponds to either unitary symmetry or antiunitary antisymmetry, while $\kappa(g)=-1$ reverses the orientation of the time-loop when $\tilde{U}_g$ corresponds to either antiunitary symmetry or unitary antisymmetry~\cite{Peng2020}.

The first cohomology group $H^1[A,M]$ for finite Abelian groups $M$ and $A$, can be computed by using the Universal coefficient theorem and the K$\ddot{\rm u}$nneth formula~\cite{Brown1982,Dodson1997,Xie_chen2013,Mesajos2013} as
\begin{align} \label{eq:Compute first cohomology:charge}
    H^1[A,M]=&\left [ H^1(A,\mathbb{Z})\otimes M\right ]\times \textrm{Tor}\left [H^2(A,\mathbb{Z}),M\right ] \nonumber \\
     =&\left [\mathbb{Z}_1 \otimes M\right ] \times \textrm{Tor}\left [A,M\right]=\textrm{Tor}\left [A,M\right]
\end{align}
where `$\times$' is the usual direct product of groups. Here, `$\otimes$' denotes the symmetric tensor product over $\mathbb{Z}$ between two Abelian groups and `$\textrm{Tor}$' represents the torsion product. From Eq.~\eqref{eq:Compute first cohomology:charge}, the group $N$ in Eq.~\eqref{eq:Identification of module N} is isomorphic to $N\cong \mathbb{Z}\times\textrm{Tor}\left [A,M\right]$, and the second cohomology group $H^2[G=M\times A,N]$ can be expressed as
\begin{align} \label{eq:Compute second cohomology:extension}
    H^2[G,N]=&\left [ H^2(G,\mathbb{Z})\otimes N\right ]\times \textrm{Tor}\left [H^3(G,\mathbb{Z}),N\right ] \nonumber \\
    =&\left [M\otimes N\right]\times \left [A\otimes N\right]\times \textrm{Tor}\left[H^2(G,U(1)),N \right] \nonumber \\
    =&\left [M\times\textrm{Tor}\left [A,M\right ] \right]\times\left [A\times\textrm{Tor}\left[A,M\right]\right] \nonumber \\
    &\times\textrm{Tor}\left[H^2[M\times A,U(1)],\textrm{Tor}\left[A,M\right]\right].
\end{align}

\begin{figure}[t]
    \centering
    \includegraphics[width=\columnwidth]{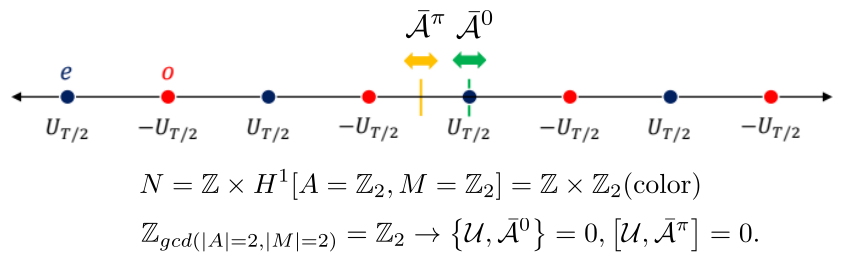}
    \caption{The frequency lattice labels each site with $n\times\Omega$ for $n\in\mathbb{Z}$, representing the discrete T-time translation of $H(t)$. Vertices of the lattice can be colored based on the interplay between the space-time symmetry $M$ and antisymmetry $A$ of the finite group $G=M\times A$. The symmetry action of $M$ onto each vertex, but applied differently to even and odd sites alternatively (the order-two of symmetry $M$), which generates a hidden color. This color is typically not observable because we can freely interchange odd and even site labels, but can be revealed by applying an additional non-local mirror operation of anisymmetry A along the lattice. The frequency lattice structure is determined by the Abelian group $N$ in Eq.~\eqref{eq:Identification of module N}, generated by the time-loop $U_F$ and colors denoted by $\bar{E}$, and is isomorphic to $\mathbb{Z}\times H^1[A=\mathbb{Z}_2,M=\mathbb{Z}_2]=\mathbb{Z}\times\mathbb{Z}_2$(color).}
    \label{fig:Decorated Frequency Lattice}
\end{figure}

We evaluate the second cohomology group for the four-step driven model $H(k,t)$ discussed in the main text. The model has the symmetry group $G=M\times A$ (isomorphic to $\mathbb{Z}_2\times \mathbb{Z}_2$) with subgroup $M=T_g$ and $A=\bar{A}_g$ representing the order-two dynamical space-time symmetry and antisymmetry respectively. We can represent $G$ as
\begin{equation} \label{eq:identification of G}
    G=\left \{ T_g^a \bar{A}_g^b |a,b\in \mathbb{Z}_2 \right \}.
\end{equation}
We extend $G$ by the group $N$ expressed as
\begin{equation} \label{eq:N as G-module in our model}
    N=\left \{ U_F(k=k^{*})^a (-1)^b|a\in \mathbb{Z}, b\in \mathbb{Z}_2 \right \}\cong\mathbb{Z}\times \mathbb{Z}_2,
\end{equation}
where $k^{*}$ denotes the reflection invariant $k$ point. From Eq.~\eqref{eq:Compute first cohomology:charge}, the group
$H^1[A,M]=\textrm{Tor}\left[\mathbb{Z}_2,\mathbb{Z}_2\right]$ is isomorphic to $\mathbb{Z}_{gcd(2,2)}=\mathbb{Z}_2$, and $\bar{E}^b=(e^{i2\pi/|H^1[A,M]|})^b=(-1)^b$ denotes the available $\mathbb{Z}_2$ colors on vertices along the frequency lattice illustrated in Fig.~\ref{fig:Decorated Frequency Lattice}. From this, we calculate the group $H^2[G,N]=\left[\mathbb{Z}_2\times\mathbb{Z}_2 \right]\times\left[\mathbb{Z}_2\times\mathbb{Z}_2 \right]\times\mathbb{Z}_2$, which characterizes the distinct classes of the extended symmetry group $\tilde{G}$ from $G$. The last $\mathbb{Z}_2$ comes from the torsion product $\textrm{Tor}\left[H^2\left[\mathbb{Z}_2\times\mathbb{Z}_2,U(1)\right],\textrm{Tor}\left[\mathbb{Z}_2,\mathbb{Z}_2\right]\right]=\mathbb{Z}_2$, as shown in Eq.~\eqref{eq:Compute second cohomology:extension}.

The symmetry $T_g$ is represented by the unitary symmetry $\tilde{U}_{T_g}$, and it acts on $N$ in Eq.~\eqref{eq:N as G-module in our model} as
\begin{equation} \label{eq:Projective representation of symmetry}
    \tilde{U}_{T_g}[U_F(k^{*})^a(-1)^b]\tilde{U}^{\dag}_{T_g}=U_F^{\kappa(T_g)\cdot a}(-1)^b=U_F^{a}(-1)^b.
\end{equation}
It preserves the orientation of the time-loop ($\kappa(T_g)=1$). The factor of the projective representation $\tilde{U}_{T_g}$ is given by   $\tilde{U}_{T_g}\tilde{U}_{T_g}=w(T_g,T_g)\tilde{U}_{\mathbb{I}}=w(T_g,T_g)=U_F^{a_{T_g}}\bar{E}^{b_{T_g}}\in N$ for $a_{T_g}\in\mathbb{Z}$ and $b_{T_g}\in\mathbb{Z}_2$. It is important to note that only the parity of $a_{T_g}$(gauge-invariant modulo $2$) represents the inequivalent classes. This can be seen by considering $\tilde{U}_{T_g}$ and $\tilde{U}^{\prime}_{T_g}=(U_F)^{m}\tilde{U}_{T_g}$ for $m\in\mathbb{Z}$, which induces the same automorphism on $N$:
\begin{align} \label{eq:parity,gauge-invariant a mod 2}
    &\tilde{U}_{T_g}U_F\tilde{U}^{\dag}_{T_g}=U_F \Leftrightarrow \tilde{U}^{\prime}_{T_g}U_F\tilde{U^{\prime}}^{\dag}_{T_g}=U_F, \nonumber \\
    &\tilde{U}_{T_g}^2=U_F^{a_{T_g}}=\left[U_F^{-m}\tilde{U}^{\prime}_{T_g}\right]^2=U_F^{-2m}\tilde{U^{\prime}}^2_{T_g}, \nonumber \\
    &\tilde{U}^2_{T_g}=U_F^{a_{T_g}}\Leftrightarrow \tilde{U^{\prime}}^2_{T_g}=U_F^{a^{\prime}_{T_g}}.
\end{align}
From the relation $a^{\prime}_{T_g}=a_{T_g}+2m$, it is indeed evident that only the parity of $a_{T_g}$ is gauge invariant. Therefore, the factor $U_F^{a_{T_g}}\bar{E}^{b_{T_g}}$ of the projective representation $\tilde{U}_{T_g}$, which is isomorphic to $\mathbb{Z}_2\times\mathbb{Z}_2$, accounts for the first extension term of $H^2[G,N]$.

The antisymmetry $\bar{A}_g$ is represented by the antiunitary antisymmetry $\tilde{U}_{\bar{A}_g}$, and it induces the automorphism on $N$ as
\begin{equation} \label{eq:Projective representatin of antisymmetry}
    \tilde{U}_{\bar{A}_g}[U_F(k^{*})^a(-1)^b]\tilde{U}^{-1}_{\bar{A}_g}=U_F^{\kappa(\bar{A}_g)\cdot a}(-1)^b=U_F^{a}(-1)^b,
\end{equation}
preserving the orientation of the time-loop $U_F$ as well. Similar to Eq.~\eqref{eq:parity,gauge-invariant a mod 2}, the factor $w(\bar{A}_g,\bar{A}_g)=U_F^{a_{\bar{A}_g}}\bar{E}^{b_{\bar{A}_g}}$, where $a_{\bar{A}_g}\in\mathbb{Z}_2$ and $b_{\bar{A}_g}\in\mathbb{Z}_2$, explains the second extension term of $H^2[G,N]$.

This $\mathbb{Z}_2$  (torsion product $\textrm{Tor}[H^3(G,\mathbb{Z}),N]$ of $H^2[G,N]$ for given $G=\mathbb{Z}_2\times\mathbb{Z}_2$) represents the commutation and anti-commutation relations between two projective representations $\tilde{U}_{T_g}$ and $\tilde{U}_{\bar{A}_g}$ denoted by $c=\pm1$ and can be written as
\begin{equation} \label{eq:torsion product of the group extension}
    \tilde{U}_{T_g}\tilde{U}_{\bar{A}_g}\tilde{U}^{\dag}_{T_g}\tilde{U}_{\bar{A}_g}^{-1}=\bar{E}^{c}.
\end{equation}
In the main text, the projective representations of the symmetry and antisymmetry of the enlarged Hamiltonian $\mathcal{H}(k)$ are expressed as $\tilde{U}_{T_g}=\mathcal{U}$ and $\tilde{U}_{\bar{A}_g}=\mathcal{\bar{A}}$ given by
\begin{equation} \label{eq:Unitary,Antiunitary Syms in Enlarged Hamiltonian}
    \mathcal{U}=\begin{pmatrix}
                \ddots &  &  & \\
                & U_{T/2} &  & \\
                &  & -U_{T/2} & \\
                &  &  & \ddots  \\
                \end{pmatrix},
    \mathcal{\bar{A}}=\begin{pmatrix}
                &  &  &  \iddots \\
                &  & \bar{A}_0 & \\
                & \bar{A}_0 &  & \\
                \iddots &  &  & \\
                \end{pmatrix},
\end{equation}
satisfying $\mathcal{U}\mathcal{H}(k)\mathcal{U}^{\dag}=\mathcal{H}(-k)$, $\mathcal{U}^2=\mathbb{I}$ and $\mathcal{\bar{A}}\mathcal{H}(k)\mathcal{\bar{A}}^{-1}=-\mathcal{H}(k),\mathcal{\bar{A}}^2=-\mathbb{I}$.

Each vertex of the frequency lattice is labeled with $n\in\mathbb{Z}$. The symmetry $\mathcal{U}$ is on-site, with alternating signs between even and odd sites as $U_{T/2}h_n(k)U^{\dag}_{T/2}=(-1)^{n}h_n(-k)$, due to the order-two nature of symmetry $T_g$. However, we cannot distinguish them with different colors since even and odd sites are interchangeable.

We notice that the additional non-local action of the antisymmetry $\mathcal{\bar{A}}$ along the lattice, with $\bar{A}_{0}h_n(k)\bar{A}^{-1}_{0}=-h_{-n}(k)$, can reveal the color on each vertex. The $\mathcal{\bar{A}}$ acts on the frequency lattice as a mirror with two possible centers: either passing through the lattice vertex or through the middle of the bond between two vertices, denoted as $\mathcal{\bar{A}}^0$ or $\mathcal{\bar{A}}^{\pi}$, respectively. Combined with the space-time symmetry $\mathcal{U}$, the two types of mirror operations $\mathcal{\bar{A}}^{0,\pi}$ can be distinguished such that $\mathcal{\bar{A}}^{\pi}$ exchanges the colors $(e\leftrightarrow o)$, whereas $\mathcal{\bar{A}}^0$ does not, as shown in Fig.~\ref{fig:Decorated Frequency Lattice}. This results in two distinct classes of group extensions: $\tilde{G}^{Tor}_0$ and $\tilde{G}^{Tor}_{\pi}$. These classes are distinguished by the different group algebras $\left[\mathcal{U},\mathcal{\bar{A}}^{0(\pi)}\right]_{+(-)}=0$ which arise from the opposite sign of $\bar{E}^c$ in Eq.~\eqref{eq:torsion product of the group extension}. This leads to different topological classification at zero($\pi$) modes as $\mathbb{Z}(\mathbb{Z}_1)$ in our model.

\subsection{Distinct topological classifications at different energy gaps}
The torsion product of the second cohomology group $H^2[G,N]$ provides a natural viewpoint to understand how a Floquet system can exhibit different topological classifications at different energy gaps in its quasienergy spectrum. The non-zero $\textrm{Tor}[A,M]$ (isomorphic to $H^1[A,M]$ for finite Abelian groups $M$ and $A$) plays a crucial role in generating these distinct classifications, as can be seen from equations \eqref{eq:Compute first cohomology:charge} and \eqref{eq:Compute second cohomology:extension}. To observe different topological classifications, it is essential to assign colors (nontrivial $\bar{E}\neq1$ in Eq.~\eqref{eq:Identification of module N}) to the lattice vertices on the frequency lattice.

This is consistent with our conventional understanding that the topological classification of a Floquet system is the same across all gaps in the quasi-energy spectrum, as long as the Altland-Zirnbauer (AZ) and crystalline symmetries are considered, extending the classification as $\mathcal{G}\in \{\mathbb{Z}_1,\mathbb{Z}_2, \mathbb{Z}\}$ to $\mathcal{G}^{\times n}$. Additionally, the presence of only space-time symmetry $M$ is not sufficient to generate distinct classifications at different energy gaps. In such cases, the group $N$ is identified as $\mathbb{Z}$, and from $H^2[G,\mathbb{Z}]=G$ for finite Abelian group $G$.

More generally, considering a one-dimensional frequency lattice and the presence of only an order-two non-local action along the lattice by the space-time antisymmetry $A$ in Floquet systems, at most $\mathbb{Z}_2$ distinct topological classifications can occur at the zero-gap and $\pi$-gap, when there coexists an even order-$D$ space-time symmetry $M$. In this case, the group $N$ is identified with $\mathbb{Z}\times\mathbb{Z}_{gcd(2,D)}$, and it can be calculated from
\begin{equation} \label{Torsion calculations_1}
    H^1[A=\mathbb{Z}_2,M=\mathbb{Z}_{D}]=\textrm{Tor}[\mathbb{Z}_2,\mathbb{Z}_{D}]=\mathbb{Z}_{gcd(2,{D})}.
\end{equation}
From Eq.~\eqref{eq:Compute second cohomology:extension}, the torsion product of $H^2[G,N]$ can be written as
\begin{align} \label{Torsion calculations_2}
    &\textrm{Tor}\left[H^2[\mathbb{Z}_D\times \mathbb{Z}_2,U(1)],\textrm{Tor}[\mathbb{Z}_2,\mathbb{Z}_D]\right] \nonumber \\
    &=\textrm{Tor}[\mathbb{Z}_{gcd(2,D)},\mathbb{Z}_{gcd(2,D)}]=\mathbb{Z}_{gcd(2,D)}.
\end{align}
This indicates that distinct topological classifications cannot occur in a Floquet system with odd order-$D$ space-time symmetry $M$.

\begin{thebibliography}{61}%
\makeatletter
\providecommand \@ifxundefined [1]{%
 \@ifx{#1\undefined}
}%
\providecommand \@ifnum [1]{%
 \ifnum #1\expandafter \@firstoftwo
 \else \expandafter \@secondoftwo
 \fi
}%
\providecommand \@ifx [1]{%
 \ifx #1\expandafter \@firstoftwo
 \else \expandafter \@secondoftwo
 \fi
}%
\providecommand \natexlab [1]{#1}%
\providecommand \enquote  [1]{``#1''}%
\providecommand \bibnamefont  [1]{#1}%
\providecommand \bibfnamefont [1]{#1}%
\providecommand \citenamefont [1]{#1}%
\providecommand \href@noop [0]{\@secondoftwo}%
\providecommand \href [0]{\begingroup \@sanitize@url \@href}%
\providecommand \@href[1]{\@@startlink{#1}\@@href}%
\providecommand \@@href[1]{\endgroup#1\@@endlink}%
\providecommand \@sanitize@url [0]{\catcode `\\12\catcode `\$12\catcode
  `\&12\catcode `\#12\catcode `\^12\catcode `\_12\catcode `\%12\relax}%
\providecommand \@@startlink[1]{}%
\providecommand \@@endlink[0]{}%
\providecommand \url  [0]{\begingroup\@sanitize@url \@url }%
\providecommand \@url [1]{\endgroup\@href {#1}{\urlprefix }}%
\providecommand \urlprefix  [0]{URL }%
\providecommand \Eprint [0]{\href }%
\providecommand \doibase [0]{http://dx.doi.org/}%
\providecommand \selectlanguage [0]{\@gobble}%
\providecommand \bibinfo  [0]{\@secondoftwo}%
\providecommand \bibfield  [0]{\@secondoftwo}%
\providecommand \translation [1]{[#1]}%
\providecommand \BibitemOpen [0]{}%
\providecommand \bibitemStop [0]{}%
\providecommand \bibitemNoStop [0]{.\EOS\space}%
\providecommand \EOS [0]{\spacefactor3000\relax}%
\providecommand \BibitemShut  [1]{\csname bibitem#1\endcsname}%
\let\auto@bib@innerbib\@empty
%</preamble>
\bibitem [{\citenamefont {Schnyder}\ \emph {et~al.}(2008)\citenamefont
  {Schnyder}, \citenamefont {Ryu}, \citenamefont {Furusaki},\ and\
  \citenamefont {Ludwig}}]{Schnyder2008}%
  \BibitemOpen
  \bibfield  {author} {\bibinfo {author} {\bibfnamefont {Andreas~P.}\
  \bibnamefont {Schnyder}}, \bibinfo {author} {\bibfnamefont {Shinsei}\
  \bibnamefont {Ryu}}, \bibinfo {author} {\bibfnamefont {Akira}\ \bibnamefont
  {Furusaki}}, \ and\ \bibinfo {author} {\bibfnamefont {Andreas W.~W.}\
  \bibnamefont {Ludwig}},\ }\bibfield  {title} {\enquote {\bibinfo {title}
  {Classification of topological insulators and superconductors in three
  spatial dimensions},}\ }\href {\doibase 10.1103/PhysRevB.78.195125}
  {\bibfield  {journal} {\bibinfo  {journal} {Phys. Rev. B}\ }\textbf {\bibinfo
  {volume} {78}},\ \bibinfo {pages} {195125} (\bibinfo {year}
  {2008})}\BibitemShut {NoStop}%
\bibitem [{\citenamefont {Kitaev}(2009)}]{Kitaev2009}%
  \BibitemOpen
  \bibfield  {author} {\bibinfo {author} {\bibfnamefont {Alexei}\ \bibnamefont
  {Kitaev}},\ }\bibfield  {title} {\enquote {\bibinfo {title} {Periodic table
  for topological insulators and superconductors},}\ }\href@noop {} {\bibfield
  {journal} {\bibinfo  {journal} {AIP Conf. Proc.}\ }\textbf {\bibinfo {volume}
  {1134}},\ \bibinfo {pages} {22} (\bibinfo {year} {2009})}\BibitemShut
  {NoStop}%
\bibitem [{\citenamefont {Ryu}\ \emph {et~al.}(2010)\citenamefont {Ryu},
  \citenamefont {Schnyder}, \citenamefont {Furusaki},\ and\ \citenamefont
  {Ludwig}}]{Ryu2010}%
  \BibitemOpen
  \bibfield  {author} {\bibinfo {author} {\bibfnamefont {Shinsei}\ \bibnamefont
  {Ryu}}, \bibinfo {author} {\bibfnamefont {Andreas~P}\ \bibnamefont
  {Schnyder}}, \bibinfo {author} {\bibfnamefont {Akira}\ \bibnamefont
  {Furusaki}}, \ and\ \bibinfo {author} {\bibfnamefont {Andreas~WW}\
  \bibnamefont {Ludwig}},\ }\bibfield  {title} {\enquote {\bibinfo {title}
  {Topological insulators and superconductors: tenfold way and dimensional
  hierarchy},}\ }\href@noop {} {\bibfield  {journal} {\bibinfo  {journal} {New
  Journal of Physics}\ }\textbf {\bibinfo {volume} {12}},\ \bibinfo {pages}
  {065010} (\bibinfo {year} {2010})}\BibitemShut {NoStop}%
\bibitem [{\citenamefont {Teo}\ and\ \citenamefont {Kane}(2010)}]{Teo2010}%
  \BibitemOpen
  \bibfield  {author} {\bibinfo {author} {\bibfnamefont {Jeffrey C.~Y.}\
  \bibnamefont {Teo}}\ and\ \bibinfo {author} {\bibfnamefont {C.~L.}\
  \bibnamefont {Kane}},\ }\bibfield  {title} {\enquote {\bibinfo {title}
  {Topological defects and gapless modes in insulators and superconductors},}\
  }\href {\doibase 10.1103/PhysRevB.82.115120} {\bibfield  {journal} {\bibinfo
  {journal} {Phys. Rev. B}\ }\textbf {\bibinfo {volume} {82}},\ \bibinfo
  {pages} {115120} (\bibinfo {year} {2010})}\BibitemShut {NoStop}%
\bibitem [{\citenamefont {Chiu}\ \emph {et~al.}(2016)\citenamefont {Chiu},
  \citenamefont {Teo}, \citenamefont {Schnyder},\ and\ \citenamefont
  {Ryu}}]{Chiu2016}%
  \BibitemOpen
  \bibfield  {author} {\bibinfo {author} {\bibfnamefont {Ching-Kai}\
  \bibnamefont {Chiu}}, \bibinfo {author} {\bibfnamefont {Jeffrey C.~Y.}\
  \bibnamefont {Teo}}, \bibinfo {author} {\bibfnamefont {Andreas~P.}\
  \bibnamefont {Schnyder}}, \ and\ \bibinfo {author} {\bibfnamefont {Shinsei}\
  \bibnamefont {Ryu}},\ }\bibfield  {title} {\enquote {\bibinfo {title}
  {Classification of topological quantum matter with symmetries},}\ }\href
  {\doibase 10.1103/RevModPhys.88.035005} {\bibfield  {journal} {\bibinfo
  {journal} {Rev. Mod. Phys.}\ }\textbf {\bibinfo {volume} {88}},\ \bibinfo
  {pages} {035005} (\bibinfo {year} {2016})}\BibitemShut {NoStop}%
\bibitem [{\citenamefont {Fu}(2011)}]{Fu2011}%
  \BibitemOpen
  \bibfield  {author} {\bibinfo {author} {\bibfnamefont {Liang}\ \bibnamefont
  {Fu}},\ }\bibfield  {title} {\enquote {\bibinfo {title} {Topological
  crystalline insulators},}\ }\href {\doibase 10.1103/PhysRevLett.106.106802}
  {\bibfield  {journal} {\bibinfo  {journal} {Phys. Rev. Lett.}\ }\textbf
  {\bibinfo {volume} {106}},\ \bibinfo {pages} {106802} (\bibinfo {year}
  {2011})}\BibitemShut {NoStop}%
\bibitem [{\citenamefont {Chiu}\ \emph {et~al.}(2013)\citenamefont {Chiu},
  \citenamefont {Yao},\ and\ \citenamefont {Ryu}}]{Chiu2013}%
  \BibitemOpen
  \bibfield  {author} {\bibinfo {author} {\bibfnamefont {Ching-Kai}\
  \bibnamefont {Chiu}}, \bibinfo {author} {\bibfnamefont {Hong}\ \bibnamefont
  {Yao}}, \ and\ \bibinfo {author} {\bibfnamefont {Shinsei}\ \bibnamefont
  {Ryu}},\ }\bibfield  {title} {\enquote {\bibinfo {title} {Classification of
  topological insulators and superconductors in the presence of reflection
  symmetry},}\ }\href {\doibase 10.1103/PhysRevB.88.075142} {\bibfield
  {journal} {\bibinfo  {journal} {Phys. Rev. B}\ }\textbf {\bibinfo {volume}
  {88}},\ \bibinfo {pages} {075142} (\bibinfo {year} {2013})}\BibitemShut
  {NoStop}%
\bibitem [{\citenamefont {Slager}\ \emph {et~al.}(2013)\citenamefont {Slager},
  \citenamefont {Mesaros}, \citenamefont {Juri{\v{c}}i{\'c}},\ and\
  \citenamefont {Zaanen}}]{Slager2013}%
  \BibitemOpen
  \bibfield  {author} {\bibinfo {author} {\bibfnamefont {Robert-Jan}\
  \bibnamefont {Slager}}, \bibinfo {author} {\bibfnamefont {Andrej}\
  \bibnamefont {Mesaros}}, \bibinfo {author} {\bibfnamefont {Vladimir}\
  \bibnamefont {Juri{\v{c}}i{\'c}}}, \ and\ \bibinfo {author} {\bibfnamefont
  {Jan}\ \bibnamefont {Zaanen}},\ }\bibfield  {title} {\enquote {\bibinfo
  {title} {The space group classification of topological band-insulators},}\
  }\href {https://www.nature.com/articles/nphys2513} {\bibfield  {journal}
  {\bibinfo  {journal} {Nature Physics}\ }\textbf {\bibinfo {volume} {9}},\
  \bibinfo {pages} {98--102} (\bibinfo {year} {2013})}\BibitemShut {NoStop}%
\bibitem [{\citenamefont {Shiozaki}\ and\ \citenamefont
  {Sato}(2014)}]{Shiozaki2014}%
  \BibitemOpen
  \bibfield  {author} {\bibinfo {author} {\bibfnamefont {Ken}\ \bibnamefont
  {Shiozaki}}\ and\ \bibinfo {author} {\bibfnamefont {Masatoshi}\ \bibnamefont
  {Sato}},\ }\bibfield  {title} {\enquote {\bibinfo {title} {Topology of
  crystalline insulators and superconductors},}\ }\href {\doibase
  10.1103/PhysRevB.90.165114} {\bibfield  {journal} {\bibinfo  {journal} {Phys.
  Rev. B}\ }\textbf {\bibinfo {volume} {90}},\ \bibinfo {pages} {165114}
  (\bibinfo {year} {2014})}\BibitemShut {NoStop}%
\bibitem [{\citenamefont {Ando}\ and\ \citenamefont {Fu}(2015)}]{Ando2015}%
  \BibitemOpen
  \bibfield  {author} {\bibinfo {author} {\bibfnamefont {Yoichi}\ \bibnamefont
  {Ando}}\ and\ \bibinfo {author} {\bibfnamefont {Liang}\ \bibnamefont {Fu}},\
  }\bibfield  {title} {\enquote {\bibinfo {title} {Topological crystalline
  insulators and topological superconductors: from concepts to materials},}\
  }\href@noop {} {\bibfield  {journal} {\bibinfo  {journal} {Annu. Rev.
  Condens. Matter Phys.}\ }\textbf {\bibinfo {volume} {6}},\ \bibinfo {pages}
  {361--381} (\bibinfo {year} {2015})}\BibitemShut {NoStop}%
\bibitem [{\citenamefont {Slager}\ \emph {et~al.}(2015)\citenamefont {Slager},
  \citenamefont {Rademaker}, \citenamefont {Zaanen},\ and\ \citenamefont
  {Balents}}]{Slager2015}%
  \BibitemOpen
  \bibfield  {author} {\bibinfo {author} {\bibfnamefont {Robert-Jan}\
  \bibnamefont {Slager}}, \bibinfo {author} {\bibfnamefont {Louk}\ \bibnamefont
  {Rademaker}}, \bibinfo {author} {\bibfnamefont {Jan}\ \bibnamefont {Zaanen}},
  \ and\ \bibinfo {author} {\bibfnamefont {Leon}\ \bibnamefont {Balents}},\
  }\bibfield  {title} {\enquote {\bibinfo {title} {Impurity-bound states and
  green's function zeros as local signatures of topology},}\ }\href {\doibase
  10.1103/PhysRevB.92.085126} {\bibfield  {journal} {\bibinfo  {journal} {Phys.
  Rev. B}\ }\textbf {\bibinfo {volume} {92}},\ \bibinfo {pages} {085126}
  (\bibinfo {year} {2015})}\BibitemShut {NoStop}%
\bibitem [{\citenamefont {Shiozaki}\ \emph {et~al.}(2016)\citenamefont
  {Shiozaki}, \citenamefont {Sato},\ and\ \citenamefont {Gomi}}]{Shiozaki2016}%
  \BibitemOpen
  \bibfield  {author} {\bibinfo {author} {\bibfnamefont {Ken}\ \bibnamefont
  {Shiozaki}}, \bibinfo {author} {\bibfnamefont {Masatoshi}\ \bibnamefont
  {Sato}}, \ and\ \bibinfo {author} {\bibfnamefont {Kiyonori}\ \bibnamefont
  {Gomi}},\ }\bibfield  {title} {\enquote {\bibinfo {title} {Topology of
  nonsymmorphic crystalline insulators and superconductors},}\ }\href {\doibase
  10.1103/PhysRevB.93.195413} {\bibfield  {journal} {\bibinfo  {journal} {Phys.
  Rev. B}\ }\textbf {\bibinfo {volume} {93}},\ \bibinfo {pages} {195413}
  (\bibinfo {year} {2016})}\BibitemShut {NoStop}%
\bibitem [{\citenamefont {Kruthoff}\ \emph {et~al.}(2017)\citenamefont
  {Kruthoff}, \citenamefont {de~Boer}, \citenamefont {van Wezel}, \citenamefont
  {Kane},\ and\ \citenamefont {Slager}}]{Kruthoff2017}%
  \BibitemOpen
  \bibfield  {author} {\bibinfo {author} {\bibfnamefont {Jorrit}\ \bibnamefont
  {Kruthoff}}, \bibinfo {author} {\bibfnamefont {Jan}\ \bibnamefont {de~Boer}},
  \bibinfo {author} {\bibfnamefont {Jasper}\ \bibnamefont {van Wezel}},
  \bibinfo {author} {\bibfnamefont {Charles~L.}\ \bibnamefont {Kane}}, \ and\
  \bibinfo {author} {\bibfnamefont {Robert-Jan}\ \bibnamefont {Slager}},\
  }\bibfield  {title} {\enquote {\bibinfo {title} {Topological classification
  of crystalline insulators through band structure combinatorics},}\ }\href
  {\doibase 10.1103/PhysRevX.7.041069} {\bibfield  {journal} {\bibinfo
  {journal} {Phys. Rev. X}\ }\textbf {\bibinfo {volume} {7}},\ \bibinfo {pages}
  {041069} (\bibinfo {year} {2017})}\BibitemShut {NoStop}%
\bibitem [{\citenamefont {Bradlyn}\ \emph {et~al.}(2017)\citenamefont
  {Bradlyn}, \citenamefont {Elcoro}, \citenamefont {Cano}, \citenamefont
  {Vergniory}, \citenamefont {Wang}, \citenamefont {Felser}, \citenamefont
  {Aroyo},\ and\ \citenamefont {Bernevig}}]{Bradlyn2017}%
  \BibitemOpen
  \bibfield  {author} {\bibinfo {author} {\bibfnamefont {Barry}\ \bibnamefont
  {Bradlyn}}, \bibinfo {author} {\bibfnamefont {L}~\bibnamefont {Elcoro}},
  \bibinfo {author} {\bibfnamefont {Jennifer}\ \bibnamefont {Cano}}, \bibinfo
  {author} {\bibfnamefont {MG}~\bibnamefont {Vergniory}}, \bibinfo {author}
  {\bibfnamefont {Zhijun}\ \bibnamefont {Wang}}, \bibinfo {author}
  {\bibfnamefont {C}~\bibnamefont {Felser}}, \bibinfo {author} {\bibfnamefont
  {Mois~I}\ \bibnamefont {Aroyo}}, \ and\ \bibinfo {author} {\bibfnamefont
  {B~Andrei}\ \bibnamefont {Bernevig}},\ }\bibfield  {title} {\enquote
  {\bibinfo {title} {Topological quantum chemistry},}\ }\href@noop {}
  {\bibfield  {journal} {\bibinfo  {journal} {Nature}\ }\textbf {\bibinfo
  {volume} {547}},\ \bibinfo {pages} {298--305} (\bibinfo {year}
  {2017})}\BibitemShut {NoStop}%
\bibitem [{\citenamefont {Song}\ \emph {et~al.}(2018)\citenamefont {Song},
  \citenamefont {Zhang}, \citenamefont {Fang},\ and\ \citenamefont
  {Fang}}]{Song2018}%
  \BibitemOpen
  \bibfield  {author} {\bibinfo {author} {\bibfnamefont {Zhida}\ \bibnamefont
  {Song}}, \bibinfo {author} {\bibfnamefont {Tiantian}\ \bibnamefont {Zhang}},
  \bibinfo {author} {\bibfnamefont {Zhong}\ \bibnamefont {Fang}}, \ and\
  \bibinfo {author} {\bibfnamefont {Chen}\ \bibnamefont {Fang}},\ }\bibfield
  {title} {\enquote {\bibinfo {title} {Quantitative mappings between symmetry
  and topology in solids},}\ }\href@noop {} {\bibfield  {journal} {\bibinfo
  {journal} {Nature communications}\ }\textbf {\bibinfo {volume} {9}},\
  \bibinfo {pages} {1--7} (\bibinfo {year} {2018})}\BibitemShut {NoStop}%
\bibitem [{\citenamefont {Song}\ \emph {et~al.}(2019)\citenamefont {Song},
  \citenamefont {Huang}, \citenamefont {Qi}, \citenamefont {Fang},\ and\
  \citenamefont {Hermele}}]{Song2019}%
  \BibitemOpen
  \bibfield  {author} {\bibinfo {author} {\bibfnamefont {Zhida}\ \bibnamefont
  {Song}}, \bibinfo {author} {\bibfnamefont {Sheng-Jie}\ \bibnamefont {Huang}},
  \bibinfo {author} {\bibfnamefont {Yang}\ \bibnamefont {Qi}}, \bibinfo
  {author} {\bibfnamefont {Chen}\ \bibnamefont {Fang}}, \ and\ \bibinfo
  {author} {\bibfnamefont {Michael}\ \bibnamefont {Hermele}},\ }\bibfield
  {title} {\enquote {\bibinfo {title} {Topological states from topological
  crystals},}\ }\href@noop {} {\bibfield  {journal} {\bibinfo  {journal}
  {Science advances}\ }\textbf {\bibinfo {volume} {5}},\ \bibinfo {pages}
  {eaax2007} (\bibinfo {year} {2019})}\BibitemShut {NoStop}%
\bibitem [{\citenamefont {Po}(2020)}]{Po2020}%
  \BibitemOpen
  \bibfield  {author} {\bibinfo {author} {\bibfnamefont {Hoi~Chun}\
  \bibnamefont {Po}},\ }\bibfield  {title} {\enquote {\bibinfo {title}
  {Symmetry indicators of band topology},}\ }\href {\doibase
  10.1088/1361-648X/ab7adb} {\bibfield  {journal} {\bibinfo  {journal} {Journal
  of Physics: Condensed Matter}\ }\textbf {\bibinfo {volume} {32}},\ \bibinfo
  {pages} {263001} (\bibinfo {year} {2020})}\BibitemShut {NoStop}%
\bibitem [{\citenamefont {Bouhon}\ \emph {et~al.}(2020)\citenamefont {Bouhon},
  \citenamefont {Wu}, \citenamefont {Slager}, \citenamefont {Weng},
  \citenamefont {Yazyev},\ and\ \citenamefont {Bzdu{\v s}ek}}]{Bouhon2020}%
  \BibitemOpen
  \bibfield  {author} {\bibinfo {author} {\bibfnamefont {Adrien}\ \bibnamefont
  {Bouhon}}, \bibinfo {author} {\bibfnamefont {QuanSheng}\ \bibnamefont {Wu}},
  \bibinfo {author} {\bibfnamefont {Robert-Jan}\ \bibnamefont {Slager}},
  \bibinfo {author} {\bibfnamefont {Hongming}\ \bibnamefont {Weng}}, \bibinfo
  {author} {\bibfnamefont {Oleg~V.}\ \bibnamefont {Yazyev}}, \ and\ \bibinfo
  {author} {\bibfnamefont {Tom{\'a}{\v s}}\ \bibnamefont {Bzdu{\v s}ek}},\
  }\bibfield  {title} {\enquote {\bibinfo {title} {Non-abelian reciprocal
  braiding of weyl points and its manifestation in zrte},}\ }\href {\doibase
  10.1038/s41567-020-0967-9} {\bibfield  {journal} {\bibinfo  {journal} {Nature
  Physics}\ }\textbf {\bibinfo {volume} {16}},\ \bibinfo {pages} {1137--1143}
  (\bibinfo {year} {2020})}\BibitemShut {NoStop}%
\bibitem [{\citenamefont {Kitagawa}\ \emph {et~al.}(2011)\citenamefont
  {Kitagawa}, \citenamefont {Oka}, \citenamefont {Brataas}, \citenamefont
  {Fu},\ and\ \citenamefont {Demler}}]{Kitagawa2011}%
  \BibitemOpen
  \bibfield  {author} {\bibinfo {author} {\bibfnamefont {Takuya}\ \bibnamefont
  {Kitagawa}}, \bibinfo {author} {\bibfnamefont {Takashi}\ \bibnamefont {Oka}},
  \bibinfo {author} {\bibfnamefont {Arne}\ \bibnamefont {Brataas}}, \bibinfo
  {author} {\bibfnamefont {Liang}\ \bibnamefont {Fu}}, \ and\ \bibinfo {author}
  {\bibfnamefont {Eugene}\ \bibnamefont {Demler}},\ }\bibfield  {title}
  {\enquote {\bibinfo {title} {Transport properties of nonequilibrium systems
  under the application of light: Photoinduced quantum hall insulators without
  landau levels},}\ }\href {\doibase 10.1103/PhysRevB.84.235108} {\bibfield
  {journal} {\bibinfo  {journal} {Phys. Rev. B}\ }\textbf {\bibinfo {volume}
  {84}},\ \bibinfo {pages} {235108} (\bibinfo {year} {2011})}\BibitemShut
  {NoStop}%
\bibitem [{\citenamefont {Rudner}\ \emph {et~al.}(2013)\citenamefont {Rudner},
  \citenamefont {Lindner}, \citenamefont {Berg},\ and\ \citenamefont
  {Levin}}]{Rudner2013}%
  \BibitemOpen
  \bibfield  {author} {\bibinfo {author} {\bibfnamefont {Mark~S.}\ \bibnamefont
  {Rudner}}, \bibinfo {author} {\bibfnamefont {Netanel~H.}\ \bibnamefont
  {Lindner}}, \bibinfo {author} {\bibfnamefont {Erez}\ \bibnamefont {Berg}}, \
  and\ \bibinfo {author} {\bibfnamefont {Michael}\ \bibnamefont {Levin}},\
  }\bibfield  {title} {\enquote {\bibinfo {title} {Anomalous edge states and
  the bulk-edge correspondence for periodically driven two-dimensional
  systems},}\ }\href {\doibase 10.1103/PhysRevX.3.031005} {\bibfield  {journal}
  {\bibinfo  {journal} {Phys. Rev. X}\ }\textbf {\bibinfo {volume} {3}},\
  \bibinfo {pages} {031005} (\bibinfo {year} {2013})}\BibitemShut {NoStop}%
\bibitem [{\citenamefont {Nathan}\ and\ \citenamefont
  {Rudner}(2015)}]{Nathan2015}%
  \BibitemOpen
  \bibfield  {author} {\bibinfo {author} {\bibfnamefont {Frederik}\
  \bibnamefont {Nathan}}\ and\ \bibinfo {author} {\bibfnamefont {Mark~S}\
  \bibnamefont {Rudner}},\ }\bibfield  {title} {\enquote {\bibinfo {title}
  {Topological singularities and the general classification of floquet--bloch
  systems},}\ }\href@noop {} {\bibfield  {journal} {\bibinfo  {journal} {New
  Journal of Physics}\ }\textbf {\bibinfo {volume} {17}},\ \bibinfo {pages}
  {125014} (\bibinfo {year} {2015})}\BibitemShut {NoStop}%
\bibitem [{\citenamefont {von Keyserlingk}\ and\ \citenamefont
  {Sondhi}(2016{\natexlab{a}})}]{Keyserlingk2016_1}%
  \BibitemOpen
  \bibfield  {author} {\bibinfo {author} {\bibfnamefont {C.~W.}\ \bibnamefont
  {von Keyserlingk}}\ and\ \bibinfo {author} {\bibfnamefont {S.~L.}\
  \bibnamefont {Sondhi}},\ }\bibfield  {title} {\enquote {\bibinfo {title}
  {Phase structure of one-dimensional interacting floquet systems. i. abelian
  symmetry-protected topological phases},}\ }\href {\doibase
  10.1103/PhysRevB.93.245145} {\bibfield  {journal} {\bibinfo  {journal} {Phys.
  Rev. B}\ }\textbf {\bibinfo {volume} {93}},\ \bibinfo {pages} {245145}
  (\bibinfo {year} {2016}{\natexlab{a}})}\BibitemShut {NoStop}%
\bibitem [{\citenamefont {von Keyserlingk}\ and\ \citenamefont
  {Sondhi}(2016{\natexlab{b}})}]{Keyserlingk2016_2}%
  \BibitemOpen
  \bibfield  {author} {\bibinfo {author} {\bibfnamefont {C.~W.}\ \bibnamefont
  {von Keyserlingk}}\ and\ \bibinfo {author} {\bibfnamefont {S.~L.}\
  \bibnamefont {Sondhi}},\ }\bibfield  {title} {\enquote {\bibinfo {title}
  {Phase structure of one-dimensional interacting floquet systems. ii.
  symmetry-broken phases},}\ }\href {\doibase 10.1103/PhysRevB.93.245146}
  {\bibfield  {journal} {\bibinfo  {journal} {Phys. Rev. B}\ }\textbf {\bibinfo
  {volume} {93}},\ \bibinfo {pages} {245146} (\bibinfo {year}
  {2016}{\natexlab{b}})}\BibitemShut {NoStop}%
\bibitem [{\citenamefont {Else}\ and\ \citenamefont {Nayak}(2016)}]{Else2016}%
  \BibitemOpen
  \bibfield  {author} {\bibinfo {author} {\bibfnamefont {Dominic~V.}\
  \bibnamefont {Else}}\ and\ \bibinfo {author} {\bibfnamefont {Chetan}\
  \bibnamefont {Nayak}},\ }\bibfield  {title} {\enquote {\bibinfo {title}
  {Classification of topological phases in periodically driven interacting
  systems},}\ }\href {\doibase 10.1103/PhysRevB.93.201103} {\bibfield
  {journal} {\bibinfo  {journal} {Phys. Rev. B}\ }\textbf {\bibinfo {volume}
  {93}},\ \bibinfo {pages} {201103} (\bibinfo {year} {2016})}\BibitemShut
  {NoStop}%
\bibitem [{\citenamefont {Potter}\ \emph {et~al.}(2016)\citenamefont {Potter},
  \citenamefont {Morimoto},\ and\ \citenamefont {Vishwanath}}]{Potter2016}%
  \BibitemOpen
  \bibfield  {author} {\bibinfo {author} {\bibfnamefont {Andrew~C.}\
  \bibnamefont {Potter}}, \bibinfo {author} {\bibfnamefont {Takahiro}\
  \bibnamefont {Morimoto}}, \ and\ \bibinfo {author} {\bibfnamefont {Ashvin}\
  \bibnamefont {Vishwanath}},\ }\bibfield  {title} {\enquote {\bibinfo {title}
  {Classification of interacting topological floquet phases in one
  dimension},}\ }\href {\doibase 10.1103/PhysRevX.6.041001} {\bibfield
  {journal} {\bibinfo  {journal} {Phys. Rev. X}\ }\textbf {\bibinfo {volume}
  {6}},\ \bibinfo {pages} {041001} (\bibinfo {year} {2016})}\BibitemShut
  {NoStop}%
\bibitem [{\citenamefont {Khemani}\ \emph {et~al.}(2016)\citenamefont
  {Khemani}, \citenamefont {Lazarides}, \citenamefont {Moessner},\ and\
  \citenamefont {Sondhi}}]{Khemani2016}%
  \BibitemOpen
  \bibfield  {author} {\bibinfo {author} {\bibfnamefont {Vedika}\ \bibnamefont
  {Khemani}}, \bibinfo {author} {\bibfnamefont {Achilleas}\ \bibnamefont
  {Lazarides}}, \bibinfo {author} {\bibfnamefont {Roderich}\ \bibnamefont
  {Moessner}}, \ and\ \bibinfo {author} {\bibfnamefont {S.~L.}\ \bibnamefont
  {Sondhi}},\ }\bibfield  {title} {\enquote {\bibinfo {title} {Phase structure
  of driven quantum systems},}\ }\href {\doibase
  10.1103/PhysRevLett.116.250401} {\bibfield  {journal} {\bibinfo  {journal}
  {Phys. Rev. Lett.}\ }\textbf {\bibinfo {volume} {116}},\ \bibinfo {pages}
  {250401} (\bibinfo {year} {2016})}\BibitemShut {NoStop}%
\bibitem [{\citenamefont {\"Unal}\ \emph {et~al.}(2019)\citenamefont {\"Unal},
  \citenamefont {Seradjeh},\ and\ \citenamefont {Eckardt}}]{Unal2019}%
  \BibitemOpen
  \bibfield  {author} {\bibinfo {author} {\bibfnamefont {F.~Nur}\ \bibnamefont
  {\"Unal}}, \bibinfo {author} {\bibfnamefont {Babak}\ \bibnamefont
  {Seradjeh}}, \ and\ \bibinfo {author} {\bibfnamefont {Andr\'e}\ \bibnamefont
  {Eckardt}},\ }\bibfield  {title} {\enquote {\bibinfo {title} {How to directly
  measure floquet topological invariants in optical lattices},}\ }\href
  {\doibase 10.1103/PhysRevLett.122.253601} {\bibfield  {journal} {\bibinfo
  {journal} {Phys. Rev. Lett.}\ }\textbf {\bibinfo {volume} {122}},\ \bibinfo
  {pages} {253601} (\bibinfo {year} {2019})}\BibitemShut {NoStop}%
\bibitem [{\citenamefont {\"Unal}\ \emph {et~al.}(2020)\citenamefont {\"Unal},
  \citenamefont {Bouhon},\ and\ \citenamefont {Slager}}]{Unal2020}%
  \BibitemOpen
  \bibfield  {author} {\bibinfo {author} {\bibfnamefont {F.~Nur}\ \bibnamefont
  {\"Unal}}, \bibinfo {author} {\bibfnamefont {Adrien}\ \bibnamefont {Bouhon}},
  \ and\ \bibinfo {author} {\bibfnamefont {Robert-Jan}\ \bibnamefont
  {Slager}},\ }\bibfield  {title} {\enquote {\bibinfo {title} {Topological
  euler class as a dynamical observable in optical lattices},}\ }\href
  {\doibase 10.1103/PhysRevLett.125.053601} {\bibfield  {journal} {\bibinfo
  {journal} {Phys. Rev. Lett.}\ }\textbf {\bibinfo {volume} {125}},\ \bibinfo
  {pages} {053601} (\bibinfo {year} {2020})}\BibitemShut {NoStop}%
\bibitem [{\citenamefont {Vu}\ \emph {et~al.}(2021)\citenamefont {Vu},
  \citenamefont {Zhang}, \citenamefont {Yang},\ and\ \citenamefont
  {Das~Sarma}}]{Vu2021}%
  \BibitemOpen
  \bibfield  {author} {\bibinfo {author} {\bibfnamefont {DinhDuy}\ \bibnamefont
  {Vu}}, \bibinfo {author} {\bibfnamefont {Rui-Xing}\ \bibnamefont {Zhang}},
  \bibinfo {author} {\bibfnamefont {Zhi-Cheng}\ \bibnamefont {Yang}}, \ and\
  \bibinfo {author} {\bibfnamefont {S.}~\bibnamefont {Das~Sarma}},\ }\bibfield
  {title} {\enquote {\bibinfo {title} {Superconductors with anomalous
  {{Floquet}} higher-order topology},}\ }\href {\doibase
  10.1103/PhysRevB.104.L140502} {\bibfield  {journal} {\bibinfo  {journal}
  {Physical Review B}\ }\textbf {\bibinfo {volume} {104}},\ \bibinfo {pages}
  {L140502} (\bibinfo {year} {2021})}\BibitemShut {NoStop}%
\bibitem [{\citenamefont {Slager}\ \emph {et~al.}(2022)\citenamefont {Slager},
  \citenamefont {Bouhon},\ and\ \citenamefont {Ünal}}]{Slager2022}%
  \BibitemOpen
  \bibfield  {author} {\bibinfo {author} {\bibfnamefont {Robert-Jan}\
  \bibnamefont {Slager}}, \bibinfo {author} {\bibfnamefont {Adrien}\
  \bibnamefont {Bouhon}}, \ and\ \bibinfo {author} {\bibfnamefont {F.~Nur}\
  \bibnamefont {Ünal}},\ }\bibfield  {title} {\enquote {\bibinfo {title}
  {Floquet multi-gap topology: Non-abelian braiding and anomalous dirac string
  phase},}\ }\href {https://arxiv.org/abs/2208.12824} {\  (\bibinfo {year}
  {2022})},\ \Eprint {http://arxiv.org/abs/2208.12824} {arXiv:2208.12824
  [cond-mat.mes-hall]} \BibitemShut {NoStop}%
\bibitem [{\citenamefont {Kitagawa}\ \emph {et~al.}(2010)\citenamefont
  {Kitagawa}, \citenamefont {Berg}, \citenamefont {Rudner},\ and\ \citenamefont
  {Demler}}]{Kitagawa2010}%
  \BibitemOpen
  \bibfield  {author} {\bibinfo {author} {\bibfnamefont {Takuya}\ \bibnamefont
  {Kitagawa}}, \bibinfo {author} {\bibfnamefont {Erez}\ \bibnamefont {Berg}},
  \bibinfo {author} {\bibfnamefont {Mark}\ \bibnamefont {Rudner}}, \ and\
  \bibinfo {author} {\bibfnamefont {Eugene}\ \bibnamefont {Demler}},\
  }\bibfield  {title} {\enquote {\bibinfo {title} {Topological characterization
  of periodically driven quantum systems},}\ }\href {\doibase
  10.1103/PhysRevB.82.235114} {\bibfield  {journal} {\bibinfo  {journal} {Phys.
  Rev. B}\ }\textbf {\bibinfo {volume} {82}},\ \bibinfo {pages} {235114}
  (\bibinfo {year} {2010})}\BibitemShut {NoStop}%
\bibitem [{\citenamefont {Titum}\ \emph {et~al.}(2016)\citenamefont {Titum},
  \citenamefont {Berg}, \citenamefont {Rudner}, \citenamefont {Refael},\ and\
  \citenamefont {Lindner}}]{Titum2016}%
  \BibitemOpen
  \bibfield  {author} {\bibinfo {author} {\bibfnamefont {Paraj}\ \bibnamefont
  {Titum}}, \bibinfo {author} {\bibfnamefont {Erez}\ \bibnamefont {Berg}},
  \bibinfo {author} {\bibfnamefont {Mark~S.}\ \bibnamefont {Rudner}}, \bibinfo
  {author} {\bibfnamefont {Gil}\ \bibnamefont {Refael}}, \ and\ \bibinfo
  {author} {\bibfnamefont {Netanel~H.}\ \bibnamefont {Lindner}},\ }\bibfield
  {title} {\enquote {\bibinfo {title} {Anomalous floquet-anderson insulator as
  a nonadiabatic quantized charge pump},}\ }\href {\doibase
  10.1103/PhysRevX.6.021013} {\bibfield  {journal} {\bibinfo  {journal} {Phys.
  Rev. X}\ }\textbf {\bibinfo {volume} {6}},\ \bibinfo {pages} {021013}
  (\bibinfo {year} {2016})}\BibitemShut {NoStop}%
\bibitem [{\citenamefont {Nathan}\ \emph {et~al.}(2017)\citenamefont {Nathan},
  \citenamefont {Rudner}, \citenamefont {Lindner}, \citenamefont {Berg},\ and\
  \citenamefont {Refael}}]{Nathan2017}%
  \BibitemOpen
  \bibfield  {author} {\bibinfo {author} {\bibfnamefont {Frederik}\
  \bibnamefont {Nathan}}, \bibinfo {author} {\bibfnamefont {Mark~S.}\
  \bibnamefont {Rudner}}, \bibinfo {author} {\bibfnamefont {Netanel~H.}\
  \bibnamefont {Lindner}}, \bibinfo {author} {\bibfnamefont {Erez}\
  \bibnamefont {Berg}}, \ and\ \bibinfo {author} {\bibfnamefont {Gil}\
  \bibnamefont {Refael}},\ }\bibfield  {title} {\enquote {\bibinfo {title}
  {Quantized magnetization density in periodically driven systems},}\ }\href
  {\doibase 10.1103/PhysRevLett.119.186801} {\bibfield  {journal} {\bibinfo
  {journal} {Phys. Rev. Lett.}\ }\textbf {\bibinfo {volume} {119}},\ \bibinfo
  {pages} {186801} (\bibinfo {year} {2017})}\BibitemShut {NoStop}%
\bibitem [{\citenamefont {Nathan}\ \emph {et~al.}(2019)\citenamefont {Nathan},
  \citenamefont {Abanin}, \citenamefont {Berg}, \citenamefont {Lindner},\ and\
  \citenamefont {Rudner}}]{Nathan2019}%
  \BibitemOpen
  \bibfield  {author} {\bibinfo {author} {\bibfnamefont {Frederik}\
  \bibnamefont {Nathan}}, \bibinfo {author} {\bibfnamefont {Dmitry}\
  \bibnamefont {Abanin}}, \bibinfo {author} {\bibfnamefont {Erez}\ \bibnamefont
  {Berg}}, \bibinfo {author} {\bibfnamefont {Netanel~H.}\ \bibnamefont
  {Lindner}}, \ and\ \bibinfo {author} {\bibfnamefont {Mark~S.}\ \bibnamefont
  {Rudner}},\ }\bibfield  {title} {\enquote {\bibinfo {title} {Anomalous
  floquet insulators},}\ }\href {\doibase 10.1103/PhysRevB.99.195133}
  {\bibfield  {journal} {\bibinfo  {journal} {Phys. Rev. B}\ }\textbf {\bibinfo
  {volume} {99}},\ \bibinfo {pages} {195133} (\bibinfo {year}
  {2019})}\BibitemShut {NoStop}%
\bibitem [{\citenamefont {Wintersperger}\ \emph {et~al.}(2020)\citenamefont
  {Wintersperger}, \citenamefont {Braun}, \citenamefont {{\"U}nal},
  \citenamefont {Eckardt}, \citenamefont {Liberto}, \citenamefont {Goldman},
  \citenamefont {Bloch},\ and\ \citenamefont
  {Aidelsburger}}]{Wintersperger2020}%
  \BibitemOpen
  \bibfield  {author} {\bibinfo {author} {\bibfnamefont {Karen}\ \bibnamefont
  {Wintersperger}}, \bibinfo {author} {\bibfnamefont {Christoph}\ \bibnamefont
  {Braun}}, \bibinfo {author} {\bibfnamefont {F.~Nur}\ \bibnamefont
  {{\"U}nal}}, \bibinfo {author} {\bibfnamefont {Andr{\'e}}\ \bibnamefont
  {Eckardt}}, \bibinfo {author} {\bibfnamefont {Marco~Di}\ \bibnamefont
  {Liberto}}, \bibinfo {author} {\bibfnamefont {Nathan}\ \bibnamefont
  {Goldman}}, \bibinfo {author} {\bibfnamefont {Immanuel}\ \bibnamefont
  {Bloch}}, \ and\ \bibinfo {author} {\bibfnamefont {Monika}\ \bibnamefont
  {Aidelsburger}},\ }\bibfield  {title} {\enquote {\bibinfo {title}
  {Realization of an anomalous floquet topological system with ultracold
  atoms},}\ }\href {\doibase 10.1038/s41567-020-0949-y} {\bibfield  {journal}
  {\bibinfo  {journal} {Nature Physics}\ }\textbf {\bibinfo {volume} {16}},\
  \bibinfo {pages} {1058--1063} (\bibinfo {year} {2020})}\BibitemShut {NoStop}%
\bibitem [{\citenamefont {Roy}\ and\ \citenamefont {Harper}(2017)}]{Roy2017}%
  \BibitemOpen
  \bibfield  {author} {\bibinfo {author} {\bibfnamefont {Rahul}\ \bibnamefont
  {Roy}}\ and\ \bibinfo {author} {\bibfnamefont {Fenner}\ \bibnamefont
  {Harper}},\ }\bibfield  {title} {\enquote {\bibinfo {title} {Periodic table
  for floquet topological insulators},}\ }\href {\doibase
  10.1103/PhysRevB.96.155118} {\bibfield  {journal} {\bibinfo  {journal} {Phys.
  Rev. B}\ }\textbf {\bibinfo {volume} {96}},\ \bibinfo {pages} {155118}
  (\bibinfo {year} {2017})}\BibitemShut {NoStop}%
\bibitem [{\citenamefont {Yao}\ \emph {et~al.}(2017)\citenamefont {Yao},
  \citenamefont {Yan},\ and\ \citenamefont {Wang}}]{Yao2017}%
  \BibitemOpen
  \bibfield  {author} {\bibinfo {author} {\bibfnamefont {Shunyu}\ \bibnamefont
  {Yao}}, \bibinfo {author} {\bibfnamefont {Zhongbo}\ \bibnamefont {Yan}}, \
  and\ \bibinfo {author} {\bibfnamefont {Zhong}\ \bibnamefont {Wang}},\
  }\bibfield  {title} {\enquote {\bibinfo {title} {Topological invariants of
  floquet systems: General formulation, special properties, and floquet
  topological defects},}\ }\href {\doibase 10.1103/PhysRevB.96.195303}
  {\bibfield  {journal} {\bibinfo  {journal} {Phys. Rev. B}\ }\textbf {\bibinfo
  {volume} {96}},\ \bibinfo {pages} {195303} (\bibinfo {year}
  {2017})}\BibitemShut {NoStop}%
\bibitem [{\citenamefont {Ladovrechis}\ and\ \citenamefont
  {Fulga}(2019)}]{Fulga2019}%
  \BibitemOpen
  \bibfield  {author} {\bibinfo {author} {\bibfnamefont {Konstantinos}\
  \bibnamefont {Ladovrechis}}\ and\ \bibinfo {author} {\bibfnamefont
  {Ion~Cosma}\ \bibnamefont {Fulga}},\ }\bibfield  {title} {\enquote {\bibinfo
  {title} {Anomalous floquet topological crystalline insulators},}\ }\href
  {\doibase 10.1103/PhysRevB.99.195426} {\bibfield  {journal} {\bibinfo
  {journal} {Phys. Rev. B}\ }\textbf {\bibinfo {volume} {99}},\ \bibinfo
  {pages} {195426} (\bibinfo {year} {2019})}\BibitemShut {NoStop}%
\bibitem [{\citenamefont {Yu}\ \emph {et~al.}(2021)\citenamefont {Yu},
  \citenamefont {Zhang},\ and\ \citenamefont {Song}}]{Yu2021}%
  \BibitemOpen
  \bibfield  {author} {\bibinfo {author} {\bibfnamefont {Jiabin}\ \bibnamefont
  {Yu}}, \bibinfo {author} {\bibfnamefont {Rui-Xing}\ \bibnamefont {Zhang}}, \
  and\ \bibinfo {author} {\bibfnamefont {Zhi-Da}\ \bibnamefont {Song}},\
  }\bibfield  {title} {\enquote {\bibinfo {title} {Dynamical symmetry
  indicators for {Floquet} crystals},}\ }\href {\doibase
  10.1038/s41467-021-26092-3} {\bibfield  {journal} {\bibinfo  {journal}
  {Nature Communications}\ }\textbf {\bibinfo {volume} {12}},\ \bibinfo {pages}
  {5985} (\bibinfo {year} {2021})}\BibitemShut {NoStop}%
\bibitem [{\citenamefont {Goldman}\ and\ \citenamefont
  {Dalibard}(2014)}]{Goldman2014}%
  \BibitemOpen
  \bibfield  {author} {\bibinfo {author} {\bibfnamefont {N.}~\bibnamefont
  {Goldman}}\ and\ \bibinfo {author} {\bibfnamefont {J.}~\bibnamefont
  {Dalibard}},\ }\bibfield  {title} {\enquote {\bibinfo {title} {Periodically
  {{Driven Quantum Systems}}: {{Effective Hamiltonians}} and {{Engineered Gauge
  Fields}}},}\ }\href {\doibase 10.1103/PhysRevX.4.031027} {\bibfield
  {journal} {\bibinfo  {journal} {Physical Review X}\ }\textbf {\bibinfo
  {volume} {4}},\ \bibinfo {pages} {031027} (\bibinfo {year}
  {2014})}\BibitemShut {NoStop}%
\bibitem [{\citenamefont {Morimoto}\ \emph {et~al.}(2017)\citenamefont
  {Morimoto}, \citenamefont {Po},\ and\ \citenamefont
  {Vishwanath}}]{Morimoto2017}%
  \BibitemOpen
  \bibfield  {author} {\bibinfo {author} {\bibfnamefont {Takahiro}\
  \bibnamefont {Morimoto}}, \bibinfo {author} {\bibfnamefont {Hoi~Chun}\
  \bibnamefont {Po}}, \ and\ \bibinfo {author} {\bibfnamefont {Ashvin}\
  \bibnamefont {Vishwanath}},\ }\bibfield  {title} {\enquote {\bibinfo {title}
  {Floquet topological phases protected by time glide symmetry},}\ }\href
  {\doibase 10.1103/PhysRevB.95.195155} {\bibfield  {journal} {\bibinfo
  {journal} {Phys. Rev. B}\ }\textbf {\bibinfo {volume} {95}},\ \bibinfo
  {pages} {195155} (\bibinfo {year} {2017})}\BibitemShut {NoStop}%
\bibitem [{\citenamefont {Xu}\ and\ \citenamefont {Wu}(2018)}]{Xu2018}%
  \BibitemOpen
  \bibfield  {author} {\bibinfo {author} {\bibfnamefont {Shenglong}\
  \bibnamefont {Xu}}\ and\ \bibinfo {author} {\bibfnamefont {Congjun}\
  \bibnamefont {Wu}},\ }\bibfield  {title} {\enquote {\bibinfo {title}
  {Space-time crystal and space-time group},}\ }\href {\doibase
  10.1103/PhysRevLett.120.096401} {\bibfield  {journal} {\bibinfo  {journal}
  {Phys. Rev. Lett.}\ }\textbf {\bibinfo {volume} {120}},\ \bibinfo {pages}
  {096401} (\bibinfo {year} {2018})}\BibitemShut {NoStop}%
\bibitem [{\citenamefont {Peng}\ and\ \citenamefont {Refael}(2019)}]{Peng2019}%
  \BibitemOpen
  \bibfield  {author} {\bibinfo {author} {\bibfnamefont {Yang}\ \bibnamefont
  {Peng}}\ and\ \bibinfo {author} {\bibfnamefont {Gil}\ \bibnamefont
  {Refael}},\ }\bibfield  {title} {\enquote {\bibinfo {title} {Floquet
  second-order topological insulators from nonsymmorphic space-time
  symmetries},}\ }\href {\doibase 10.1103/PhysRevLett.123.016806} {\bibfield
  {journal} {\bibinfo  {journal} {Phys. Rev. Lett.}\ }\textbf {\bibinfo
  {volume} {123}},\ \bibinfo {pages} {016806} (\bibinfo {year}
  {2019})}\BibitemShut {NoStop}%
\bibitem [{\citenamefont {Peng}(2020)}]{Peng2020}%
  \BibitemOpen
  \bibfield  {author} {\bibinfo {author} {\bibfnamefont {Yang}\ \bibnamefont
  {Peng}},\ }\bibfield  {title} {\enquote {\bibinfo {title} {Floquet
  higher-order topological insulators and superconductors with space-time
  symmetries},}\ }\href {\doibase 10.1103/PhysRevResearch.2.013124} {\bibfield
  {journal} {\bibinfo  {journal} {Phys. Rev. Research}\ }\textbf {\bibinfo
  {volume} {2}},\ \bibinfo {pages} {013124} (\bibinfo {year}
  {2020})}\BibitemShut {NoStop}%
\bibitem [{\citenamefont {Chaudhary}\ \emph {et~al.}(2020)\citenamefont
  {Chaudhary}, \citenamefont {Haim}, \citenamefont {Peng},\ and\ \citenamefont
  {Refael}}]{Swati2020}%
  \BibitemOpen
  \bibfield  {author} {\bibinfo {author} {\bibfnamefont {Swati}\ \bibnamefont
  {Chaudhary}}, \bibinfo {author} {\bibfnamefont {Arbel}\ \bibnamefont {Haim}},
  \bibinfo {author} {\bibfnamefont {Yang}\ \bibnamefont {Peng}}, \ and\
  \bibinfo {author} {\bibfnamefont {Gil}\ \bibnamefont {Refael}},\ }\bibfield
  {title} {\enquote {\bibinfo {title} {Phonon-induced floquet topological
  phases protected by space-time symmetries},}\ }\href {\doibase
  10.1103/PhysRevResearch.2.043431} {\bibfield  {journal} {\bibinfo  {journal}
  {Phys. Rev. Res.}\ }\textbf {\bibinfo {volume} {2}},\ \bibinfo {pages}
  {043431} (\bibinfo {year} {2020})}\BibitemShut {NoStop}%
\bibitem [{\citenamefont {Peng}(2022)}]{Peng2022}%
  \BibitemOpen
  \bibfield  {author} {\bibinfo {author} {\bibfnamefont {Yang}\ \bibnamefont
  {Peng}},\ }\bibfield  {title} {\enquote {\bibinfo {title} {Topological
  space-time crystal},}\ }\href {\doibase 10.1103/PhysRevLett.128.186802}
  {\bibfield  {journal} {\bibinfo  {journal} {Phys. Rev. Lett.}\ }\textbf
  {\bibinfo {volume} {128}},\ \bibinfo {pages} {186802} (\bibinfo {year}
  {2022})}\BibitemShut {NoStop}%
\bibitem [{SM()}]{SM}%
  \BibitemOpen
  \href@noop {} {\ }\bibinfo {note} {See Supplementary material for further
  details.}\BibitemShut {Stop}%
\bibitem [{\citenamefont {Bukov}\ \emph {et~al.}(2015)\citenamefont {Bukov},
  \citenamefont {D'Alessio},\ and\ \citenamefont
  {Polkovnikov}}]{MarinBukov2015}%
  \BibitemOpen
  \bibfield  {author} {\bibinfo {author} {\bibfnamefont {Marin}\ \bibnamefont
  {Bukov}}, \bibinfo {author} {\bibfnamefont {Luca}\ \bibnamefont {D'Alessio}},
  \ and\ \bibinfo {author} {\bibfnamefont {Anatoli}\ \bibnamefont
  {Polkovnikov}},\ }\bibfield  {title} {\enquote {\bibinfo {title} {Universal
  high-frequency behavior of periodically driven systems: from dynamical
  stabilization to floquet engineering},}\ }\href {\doibase
  10.1080/00018732.2015.1055918} {\bibfield  {journal} {\bibinfo  {journal}
  {Advances in Physics}\ }\textbf {\bibinfo {volume} {64}},\ \bibinfo {pages}
  {139--226} (\bibinfo {year} {2015})},\ \Eprint
  {http://arxiv.org/abs/https://doi.org/10.1080/00018732.2015.1055918}
  {https://doi.org/10.1080/00018732.2015.1055918} \BibitemShut {NoStop}%
\bibitem [{\citenamefont {Mesaros}\ and\ \citenamefont
  {Ran}(2013)}]{Mesajos2013}%
  \BibitemOpen
  \bibfield  {author} {\bibinfo {author} {\bibfnamefont {Andrej}\ \bibnamefont
  {Mesaros}}\ and\ \bibinfo {author} {\bibfnamefont {Ying}\ \bibnamefont
  {Ran}},\ }\bibfield  {title} {\enquote {\bibinfo {title} {Classification of
  symmetry enriched topological phases with exactly solvable models},}\ }\href
  {\doibase 10.1103/PhysRevB.87.155115} {\bibfield  {journal} {\bibinfo
  {journal} {Phys. Rev. B}\ }\textbf {\bibinfo {volume} {87}},\ \bibinfo
  {pages} {155115} (\bibinfo {year} {2013})}\BibitemShut {NoStop}%
\bibitem [{\citenamefont {Martin}\ \emph {et~al.}(2017)\citenamefont {Martin},
  \citenamefont {Refael},\ and\ \citenamefont {Halperin}}]{Martin2017}%
  \BibitemOpen
  \bibfield  {author} {\bibinfo {author} {\bibfnamefont {Ivar}\ \bibnamefont
  {Martin}}, \bibinfo {author} {\bibfnamefont {Gil}\ \bibnamefont {Refael}}, \
  and\ \bibinfo {author} {\bibfnamefont {Bertrand}\ \bibnamefont {Halperin}},\
  }\bibfield  {title} {\enquote {\bibinfo {title} {Topological frequency
  conversion in strongly driven quantum systems},}\ }\href {\doibase
  10.1103/PhysRevX.7.041008} {\bibfield  {journal} {\bibinfo  {journal} {Phys.
  Rev. X}\ }\textbf {\bibinfo {volume} {7}},\ \bibinfo {pages} {041008}
  (\bibinfo {year} {2017})}\BibitemShut {NoStop}%
\bibitem [{\citenamefont {Peng}\ and\ \citenamefont
  {Refael}(2018{\natexlab{a}})}]{Yang2018}%
  \BibitemOpen
  \bibfield  {author} {\bibinfo {author} {\bibfnamefont {Yang}\ \bibnamefont
  {Peng}}\ and\ \bibinfo {author} {\bibfnamefont {Gil}\ \bibnamefont
  {Refael}},\ }\bibfield  {title} {\enquote {\bibinfo {title}
  {Time-quasiperiodic topological superconductors with majorana
  multiplexing},}\ }\href {\doibase 10.1103/PhysRevB.98.220509} {\bibfield
  {journal} {\bibinfo  {journal} {Phys. Rev. B}\ }\textbf {\bibinfo {volume}
  {98}},\ \bibinfo {pages} {220509} (\bibinfo {year}
  {2018}{\natexlab{a}})}\BibitemShut {NoStop}%
\bibitem [{\citenamefont {Peng}\ and\ \citenamefont
  {Refael}(2018{\natexlab{b}})}]{Peng2018}%
  \BibitemOpen
  \bibfield  {author} {\bibinfo {author} {\bibfnamefont {Yang}\ \bibnamefont
  {Peng}}\ and\ \bibinfo {author} {\bibfnamefont {Gil}\ \bibnamefont
  {Refael}},\ }\bibfield  {title} {\enquote {\bibinfo {title} {Topological
  energy conversion through the bulk or the boundary of driven systems},}\
  }\href {\doibase 10.1103/PhysRevB.97.134303} {\bibfield  {journal} {\bibinfo
  {journal} {Phys. Rev. B}\ }\textbf {\bibinfo {volume} {97}},\ \bibinfo
  {pages} {134303} (\bibinfo {year} {2018}{\natexlab{b}})}\BibitemShut
  {NoStop}%
\bibitem [{\citenamefont {Crowley}\ \emph {et~al.}(2019)\citenamefont
  {Crowley}, \citenamefont {Martin},\ and\ \citenamefont
  {Chandran}}]{Crowley2019}%
  \BibitemOpen
  \bibfield  {author} {\bibinfo {author} {\bibfnamefont {P.~J.~D.}\
  \bibnamefont {Crowley}}, \bibinfo {author} {\bibfnamefont {I.}~\bibnamefont
  {Martin}}, \ and\ \bibinfo {author} {\bibfnamefont {A.}~\bibnamefont
  {Chandran}},\ }\bibfield  {title} {\enquote {\bibinfo {title} {Topological
  classification of quasiperiodically driven quantum systems},}\ }\href
  {\doibase 10.1103/PhysRevB.99.064306} {\bibfield  {journal} {\bibinfo
  {journal} {Phys. Rev. B}\ }\textbf {\bibinfo {volume} {99}},\ \bibinfo
  {pages} {064306} (\bibinfo {year} {2019})}\BibitemShut {NoStop}%
\bibitem [{\citenamefont {Potirniche}\ \emph {et~al.}(2017)\citenamefont
  {Potirniche}, \citenamefont {Potter}, \citenamefont {Schleier-Smith},
  \citenamefont {Vishwanath},\ and\ \citenamefont {Yao}}]{Potirniche2017}%
  \BibitemOpen
  \bibfield  {author} {\bibinfo {author} {\bibfnamefont {I.-D.}\ \bibnamefont
  {Potirniche}}, \bibinfo {author} {\bibfnamefont {A.~C.}\ \bibnamefont
  {Potter}}, \bibinfo {author} {\bibfnamefont {M.}~\bibnamefont
  {Schleier-Smith}}, \bibinfo {author} {\bibfnamefont {A.}~\bibnamefont
  {Vishwanath}}, \ and\ \bibinfo {author} {\bibfnamefont {N.~Y.}\ \bibnamefont
  {Yao}},\ }\bibfield  {title} {\enquote {\bibinfo {title} {Floquet
  symmetry-protected topological phases in cold-atom systems},}\ }\href
  {\doibase 10.1103/PhysRevLett.119.123601} {\bibfield  {journal} {\bibinfo
  {journal} {Phys. Rev. Lett.}\ }\textbf {\bibinfo {volume} {119}},\ \bibinfo
  {pages} {123601} (\bibinfo {year} {2017})}\BibitemShut {NoStop}%
\bibitem [{\citenamefont {Else}\ \emph {et~al.}(2017)\citenamefont {Else},
  \citenamefont {Fendley}, \citenamefont {Kemp},\ and\ \citenamefont
  {Nayak}}]{Else2017}%
  \BibitemOpen
  \bibfield  {author} {\bibinfo {author} {\bibfnamefont {Dominic~V.}\
  \bibnamefont {Else}}, \bibinfo {author} {\bibfnamefont {Paul}\ \bibnamefont
  {Fendley}}, \bibinfo {author} {\bibfnamefont {Jack}\ \bibnamefont {Kemp}}, \
  and\ \bibinfo {author} {\bibfnamefont {Chetan}\ \bibnamefont {Nayak}},\
  }\bibfield  {title} {\enquote {\bibinfo {title} {Prethermal strong zero modes
  and topological qubits},}\ }\href {\doibase 10.1103/PhysRevX.7.041062}
  {\bibfield  {journal} {\bibinfo  {journal} {Phys. Rev. X}\ }\textbf {\bibinfo
  {volume} {7}},\ \bibinfo {pages} {041062} (\bibinfo {year}
  {2017})}\BibitemShut {NoStop}%
\bibitem [{\citenamefont {Fu}\ \emph {et~al.}(2007)\citenamefont {Fu},
  \citenamefont {Kane},\ and\ \citenamefont {Mele}}]{Fu2007}%
  \BibitemOpen
  \bibfield  {author} {\bibinfo {author} {\bibfnamefont {Liang}\ \bibnamefont
  {Fu}}, \bibinfo {author} {\bibfnamefont {C.~L.}\ \bibnamefont {Kane}}, \ and\
  \bibinfo {author} {\bibfnamefont {E.~J.}\ \bibnamefont {Mele}},\ }\bibfield
  {title} {\enquote {\bibinfo {title} {Topological insulators in three
  dimensions},}\ }\href {\doibase 10.1103/PhysRevLett.98.106803} {\bibfield
  {journal} {\bibinfo  {journal} {Phys. Rev. Lett.}\ }\textbf {\bibinfo
  {volume} {98}},\ \bibinfo {pages} {106803} (\bibinfo {year}
  {2007})}\BibitemShut {NoStop}%
\bibitem [{\citenamefont {Chen}\ \emph {et~al.}(2014)\citenamefont {Chen},
  \citenamefont {Lu},\ and\ \citenamefont {Vishwanath}}]{Chen2014}%
  \BibitemOpen
  \bibfield  {author} {\bibinfo {author} {\bibfnamefont {Xie}\ \bibnamefont
  {Chen}}, \bibinfo {author} {\bibfnamefont {Yuan-Ming}\ \bibnamefont {Lu}}, \
  and\ \bibinfo {author} {\bibfnamefont {Ashvin}\ \bibnamefont {Vishwanath}},\
  }\bibfield  {title} {\enquote {\bibinfo {title} {Symmetry-protected
  topological phases from decorated domain walls},}\ }\href {\doibase
  10.1038/ncomms4507} {\bibfield  {journal} {\bibinfo  {journal} {Nature
  Communications}\ }\textbf {\bibinfo {volume} {5}},\ \bibinfo {pages} {3507}
  (\bibinfo {year} {2014})}\BibitemShut {NoStop}%
\bibitem [{\citenamefont {Chen}\ \emph {et~al.}(2013)\citenamefont {Chen},
  \citenamefont {Gu}, \citenamefont {Liu},\ and\ \citenamefont
  {Wen}}]{Xie_chen2013}%
  \BibitemOpen
  \bibfield  {author} {\bibinfo {author} {\bibfnamefont {Xie}\ \bibnamefont
  {Chen}}, \bibinfo {author} {\bibfnamefont {Zheng-Cheng}\ \bibnamefont {Gu}},
  \bibinfo {author} {\bibfnamefont {Zheng-Xin}\ \bibnamefont {Liu}}, \ and\
  \bibinfo {author} {\bibfnamefont {Xiao-Gang}\ \bibnamefont {Wen}},\
  }\bibfield  {title} {\enquote {\bibinfo {title} {Symmetry protected
  topological orders and the group cohomology of their symmetry group},}\
  }\href {\doibase 10.1103/PhysRevB.87.155114} {\bibfield  {journal} {\bibinfo
  {journal} {Phys. Rev. B}\ }\textbf {\bibinfo {volume} {87}},\ \bibinfo
  {pages} {155114} (\bibinfo {year} {2013})}\BibitemShut {NoStop}%
\bibitem [{\citenamefont {Alexandradinata}\ \emph {et~al.}(2016)\citenamefont
  {Alexandradinata}, \citenamefont {Wang},\ and\ \citenamefont
  {Bernevig}}]{Alexandradinata2016}%
  \BibitemOpen
  \bibfield  {author} {\bibinfo {author} {\bibfnamefont {A.}~\bibnamefont
  {Alexandradinata}}, \bibinfo {author} {\bibfnamefont {Zhijun}\ \bibnamefont
  {Wang}}, \ and\ \bibinfo {author} {\bibfnamefont {B.~Andrei}\ \bibnamefont
  {Bernevig}},\ }\bibfield  {title} {\enquote {\bibinfo {title} {Topological
  insulators from group cohomology},}\ }\href {\doibase
  10.1103/PhysRevX.6.021008} {\bibfield  {journal} {\bibinfo  {journal} {Phys.
  Rev. X}\ }\textbf {\bibinfo {volume} {6}},\ \bibinfo {pages} {021008}
  (\bibinfo {year} {2016})}\BibitemShut {NoStop}%
\bibitem [{\citenamefont {Brown}(1982)}]{Brown1982}%
  \BibitemOpen
  \bibfield  {author} {\bibinfo {author} {\bibfnamefont {Kenneth~S}\
  \bibnamefont {Brown}},\ }\href@noop {} {\emph {\bibinfo {title} {Cohomology
  of Groups}}},\ \bibinfo {edition} {1st}\ ed.,\ Graduate Texts in Mathematics\
  (\bibinfo  {publisher} {Springer},\ \bibinfo {address} {New York, NY},\
  \bibinfo {year} {1982})\BibitemShut {NoStop}%
\bibitem [{\citenamefont {Dodson}\ and\ \citenamefont
  {Parker}(1996)}]{Dodson1997}%
  \BibitemOpen
  \bibfield  {author} {\bibinfo {author} {\bibfnamefont {Christopher T~J}\
  \bibnamefont {Dodson}}\ and\ \bibinfo {author} {\bibfnamefont {Phillip~E}\
  \bibnamefont {Parker}},\ }\href@noop {} {\emph {\bibinfo {title} {A user's
  guide to algebraic topology}}},\ \bibinfo {edition} {1997th}\ ed.,\
  Mathematics and Its Applications\ (\bibinfo  {publisher} {Springer},\
  \bibinfo {address} {Dordrecht, Netherlands},\ \bibinfo {year}
  {1996})\BibitemShut {NoStop}%
\end{thebibliography}
\end{document}